\documentclass[aps,prb,twocolumn,superscriptaddress]{revtex4-1}

\usepackage{graphicx}
\usepackage{epsfig}
\usepackage{epstopdf}
\usepackage{float}
\usepackage{subfigure}
\usepackage{amsmath}

\usepackage{needspace}

\begin{document}

%\preprint{}

\title{Defect ordering and defect-domain wall interactions in PbTiO$_3$: A first-principles study}


\author{Anand Chandrasekaran}
\affiliation{Theory and Simulation of Materials,\'Ecole Polytechnique F\'ed\'erale de Lausanne, 1015 Lausanne, Switzerland}
\affiliation{Ceramics Laboratory, \'Ecole Polytechnique F\'ed\'erale de Lausanne, 1015 Lausanne, Switzerland}
\author{Dragan Damjanovic}
\author{Nava Setter}
\affiliation{Ceramics Laboratory, \'Ecole Polytechnique F\'ed\'erale de Lausanne, 1015 Lausanne, Switzerland}
\author{Nicola Marzari}
\affiliation{Theory and Simulation of Materials,\'Ecole Polytechnique F\'ed\'erale de Lausanne, 1015 Lausanne, Switzerland}


\date{\today}

\begin{abstract}
The properties of ferroelectric materials, such as lead zirconate
titanate (PZT), are heavily influenced by the interaction of defects
with domain walls. These defects are either intrinsic, or are induced by
the addition of dopants. We study here PbTiO$_3$ (the end member of a
key family of solid solutions) in the presence of acceptor (Fe) and
donor (Nb) dopants, and the interactions of the different defects and
defect associates with the domain walls. For the case iron acceptors,
the calculations point to the formation of defect associates involving
an iron substitutional defect and a charged oxygen vacancy
(Fe$^{'}_{Ti}$-V$^{^{\textbf{..}}}_O$). This associate exhibits a strong
tendency to align in the direction of the bulk polarization; in fact,
ordering of defects is also observed in pure PbTiO$_3$ in the form of
lead-oxygen divacancies. Conversely, calculations on donor-doped
PbTiO$_3$ do not indicate the formation of polar defect complexes
involving donor substitutions. Last, it is observed that both isolated
defects in donor-doped materials and defect associates in acceptor-doped
materials are more stable at 180$^o$ domain walls.  However, polar
defect complexes lead to asymmetric potentials at domain walls due to
the interaction of the defect polarization with the bulk polarization.
The relative pinning characteristics of different defects are then
compared, to develop an understanding of defect-domain wall interactions
in both doped and pure PbTiO$_3$. These results may also help
understanding hardening and softening mechanisms in PZT.
\end{abstract}
%\keywords{}

\maketitle

\section{Introduction }
Hardening and softening of ferroelectric materials through the addition of dopants is a key technique to tailor their properties. The best known examples are hard and soft Pb(Zr,Ti)O$_3$ (or PZT) ceramics, the most widely used piezoelectric materials. Hardening can be caused by the addition of acceptor dopants and softening by the addition of donor dopants \cite{Jaffe1971}. Hard ferroelectric materials exhibit strong aging (see later), a pinched hysteresis loop, lower electromechanical coupling coefficients, low dielectric losses, and moderate conductivity. Soft materials on the other hand exhibit large electromechanical coefficients, a square hysteresis loop, weak aging, low conductivity, and high dielectric losses. While the mechanisms of softening are still not clearly understood, there is a relatively better understanding of the phenomenon of hardening. The properties of acceptor-doped hard materials have been attributed to inhibited domain wall movement,\cite{Lambeck1986,Zhang2005,Genenko2007,Genenko2008,Zhang2005,Feng2008,Lambeck1986,Keve1972,Robels1993,Zhang2008a,Warren1996} whereas softening is thought to be associated with highly mobile domain walls. On the other hand, the properties of pure undoped PZT more closely resemble hard PZT rather than soft and the reason for this is not yet clear. The mobility of domain walls thus depends strongly on their interaction with different defects that are either present intrinsically in the material or are induced by the addition of dopants.

It has been proposed soon after discovery of hard and soft piezoelectrics\cite{Carl1977}and confirmed recently by electron paramagnetic resonance\cite{Mestric2004} and ab-initio calculations\cite{Erhart2007,Erhart2013} that in the case of acceptor-doped materials a defect associate is formed between an acceptor substitutional defect and an oxygen vacancy. It has been suggested that these defect associates align in the direction of the lattice polarization and act as pinning centers to inhibit domain wall movement. While this accounts for the aging process in hard materials, it does not explain its absence in donor-doped compositions where defect complexes also occur but are supposed to be formed between lead vacancy and donor substitutional defect.

More generally, to our knowledge there have been no calculations to provide an atomistic insight into the interaction of defect complexes with domain walls. It is not clear why soft materials have higher domain wall mobilities compared to those of undoped materials, and several hypothesis have been suggested. For instance, donor dopants are thought to compensate the effects of acceptor cations or lead vacancies that are naturally present in the undoped materials \cite{Jaffe1971}, thus preventing the formation of oxygen vacancies which are suspected to be responsible for pinning domain walls \cite{He2003}. The possibility that lead vacancies reduce internal stresses in ceramics and make domain walls more mobile has also been suggested \cite{Gerson1960}. Last, there is a possibility that electron transfer between defects could minimize the space charge at domain walls, thereby increasing domain wall mobility\cite{Eyraud2006}.

This paper investigates the nature of defects/defect associates in acceptor-doped, pure and donor-doped PbTiO$_3$ and also shows how these entities interact with 180$^o$ domain walls. Lead titanate is chosen because it is an end member of the most important family of piezoelectric and ferroelectric solid solutions (e.g., PZT and Pb(Mg$_{1/3}$Nb$_{2/3}$O$_3$-PbTiO$_3$)\cite{Safari2008} and doped PbTiO$_3$ itself is employed in some applications\cite{Whatmore1986,Takeuchi1982} . While Pb(Zr,Ti)O$_3$ is structurally more complex near its morphotropic phase boundary, it is expected to behave as PbTiO$_3$ on the tetragonal side of the phase diagram. Although a lot of work has been done on the preferential alignment of metal-oxygen vacancy complexes\cite{Erhart2013,Erhart2007,Meyer2002}, it is not yet clear whether lead-oxygen divacancies (presumed to be the most common defects in pure lead titanate and in PZT)  display similar behavior. Moreover, calculations of defect associates involving donor substitutional defects have also never been reported to our knowledge.

Our calculations indicate that both pure PbTiO$_3$ and acceptor-doped PbTiO$_3$ exhibit ordered defect complexes which are aligned with the polarization in contrast to donor-doped materials where association of defects is found to be weaker and alignment with polarization absent. Based on these results, we explain the barrier energies and potential energy landscape of domain walls in the vicinity of these defects and defect associates. Absence of ordering for the defect complexes in donor-doped materials and the lower energy barriers for domain walls motion thus rationalize the absence of ageing and weaker pinning than in acceptor-doped materials. 

The paper is organized as follows: In Sec.II we describe the computational details of our first-principles calculations. In Sec.III the different configurations of ordered defect associates in pure and Fe-doped PbTiO$_3$ are studied and our results are discussed in the context of previous computational and experimental work. In Sec.IV, we study the nature of defect complexes in Nb donor-doped PbTiO$_3$. In Sec.V, we present our results on defect-domain wall interactions and finally the paper concludes with a summary in Sec.VI. We note that some calculations on alignment of defects complexes in acceptor-doped materials in Section III are similar to those recently published by Erhart et al. \cite{Erhart2013},but are also independently reported here for consistency.

\section{Computational Details}
We used density-functional theory in the local-density approximation using ultrasoft pseudopotentials and plane waves, as implemented in the Quantum ESPRESSO distribution\cite{Giannozzi2009}. Although the ultimate material of interest is PZT, all calculations are performed on the ground-state of lead titanate (one of the end members of PZT phase diagram), which has a tetragonal structure. First-principles calculations have shown, for example, that Pb(Zr$_{0.5}$Ti$_{0.5}$)O$_3$ and PbTiO$_3$ do not display significant differences in their local atomic structure and they have similar spontaneous polarization, dynamical charges and piezoelectric moduli\cite{Saghi-Szabo1999}. Hence, the results of this paper can be deemed relevant for a broader class of materials including PZT. For the defect calculations, ground-state energy calculations are performed on 3*3*3 supercells containing 135 atoms. A 2*2*2 Monkhorst-Pack \emph{k}-point mesh, a plane wave cutoff of 40 Ry and a charge density cutoff 240 Ry  are utilized. The plane wave cut-off is chosen to converge the domain wall formation energy to 1 meV.  The calculations are spin polarized in all supercells containing iron. The atomic positions are allowed to relax until forces on atoms are smaller than $10^{-4}$ Ry/bohr. In the case of supercells containing a charged defect, a compensating jellium background of opposite charge is inserted to remove divergencies.


\section{Ordered defect complexes}
For the calculation of the acceptor oxygen vacancy defect associate, one Ti atom in the 3*3*3 supercell was replaced by a Fe atom and the stability of oxygen vacancies at different positions with respect to the Fe atom was investigated. EPR experiments have shown that iron is in the trivalent state\cite{Mestric2004} and hence the substitutional defect is negatively charged. Recently, first-principles calculations have also confirmed that oxygen vacancies are doubly (positively) charged\cite{Yao2011}. Thus, calculations on supercells containing the Fe$^{'}_{Ti}$-V$^{^{\textbf{..}}}_O$ defect associate (Kr\"{o}ger-Vink notation, where prime symbol denotes -1 charge and dot +1 charge) are performed with a net positive charge +1.   {Fig.\ref{fig:1} shows the axial configuration of the Fe$^{'}_{Ti}$-V$^{^{\textbf{..}}}_O$ defect associate along the direction of polarization, Fig.\ref{fig:2} shows the equatorial configuration and Fig.\ref{fig:3} anti-axial configuration
The lowest energy configuration is the one in which the defect associate is oriented in the direction of polarization. The equatorial configuration of the Fe$^{'}_{Ti}$-V$^{^{\textbf{..}}}_O$ defect associate has a higher energy than the antiaxial state, suggesting that elastic effects are also important and the highest energy configuration is the one in which the two defects are separated from each other (not shown). These results shows that the two defects indeed are driven to form a defect associate rather than remaining isolated. All these calculations are in agreement with those performed by Erhart et al.\cite{Erhart2007}, and with both the bulk stabilization effect of defect associates and the symmetry conforming mechanisms \cite{Zhang2005,Nowick1965}.

Aging is then related to the time it takes for oxygen vacancies to hop from randomly oriented  configurations to the aligned configuration below the Curie temperature, and deaging is the electromechanical process required to reverse this alignment process. Morozov et al. \cite{Morozov2008} observed that the  AC conductivity for iron-doped PZT is a temperature-dependent Arrhenius process. The activation energy for this process was experimentally determined to be between 0.6 and 0.8 eV, in agreement with what was observed in deaging.   

In order to verify if the energetics for oxygen vacancy migration matches that of conductivity and ageing we calculated all relevant barriers using nudged elastic-band calculations\cite{Henkelman2000}, and a minimum energy pathway was identified for the diffusion of the oxygen vacancy from the aligned state (Fig.\ref{fig:1}) to the equatorial state (Fig.\ref{fig:2}). Fig.\ref{nebFeVO} shows the potential energy surface for this process, with an activation energy of 0.89 eV in good agreement with Morozov et al. \cite{Morozov2008}. This value is slightly higher than the the barrier energy calculated very recently by Erhart et al.\cite{Erhart2013} (0.84 eV); the small discrepancies likely due to differences in pseudopotentials utilized.

 \begin{figure*}[!ht]
\centering
\subfigure[The axial ground-state structure, with the Fe$^{'}_{Ti}$-V$^{^{\textbf{..}}}_O$ defect associate oriented in the direction of lattice polarization. Magnetization is 3 Bohr mag/cell]
{
 \includegraphics[width=0.315\textwidth]{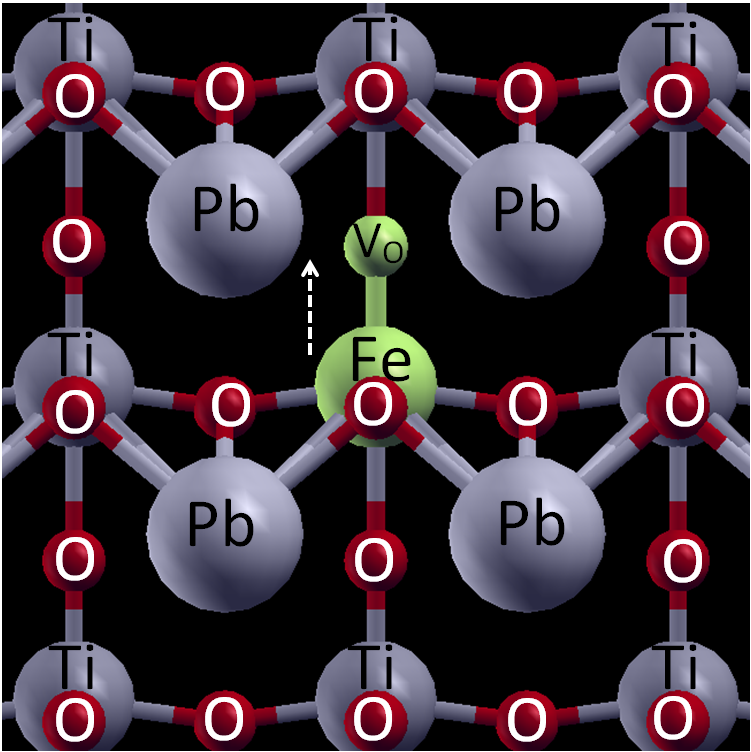}
  \label{fig:1}
}
\subfigure[In the equatorial state, the defect associate is oriented perpendicular to the lattice polarization. The energy of this configuration is 0.52 eV higher than the ground-state structure. Magnetization is 5 Bohr mag/cell]
{
 \includegraphics[width=0.315\textwidth]{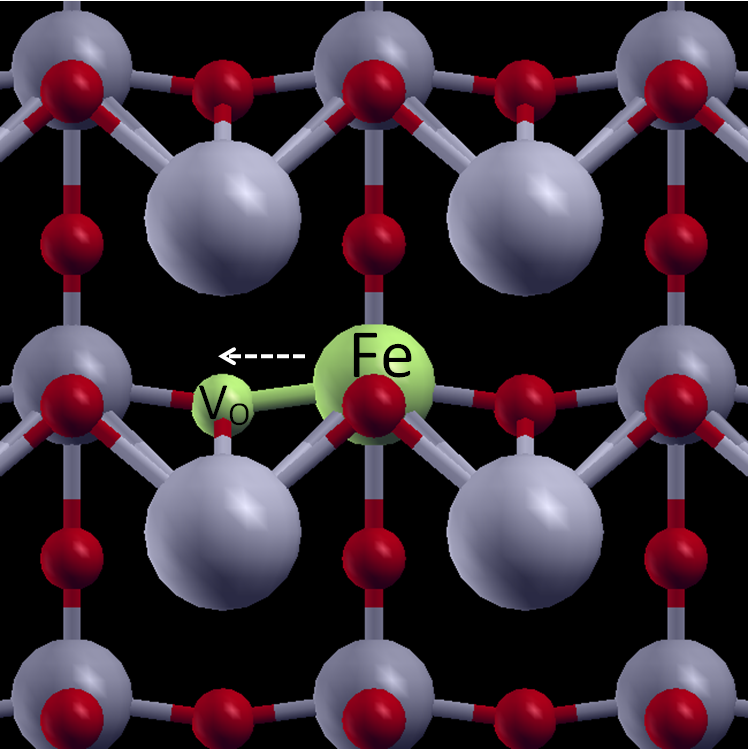}
  \label{fig:2}
}
\subfigure[The anti-axial state with Fe$^{'}_{Ti}$-V$^{^{\textbf{..}}}_O$ oriented in the opposite direction has an energy of 0.38 eV higher than the ground-state. Magnetization is 3 Bohr mag/cell]
{
 \includegraphics[width=0.305\textwidth]{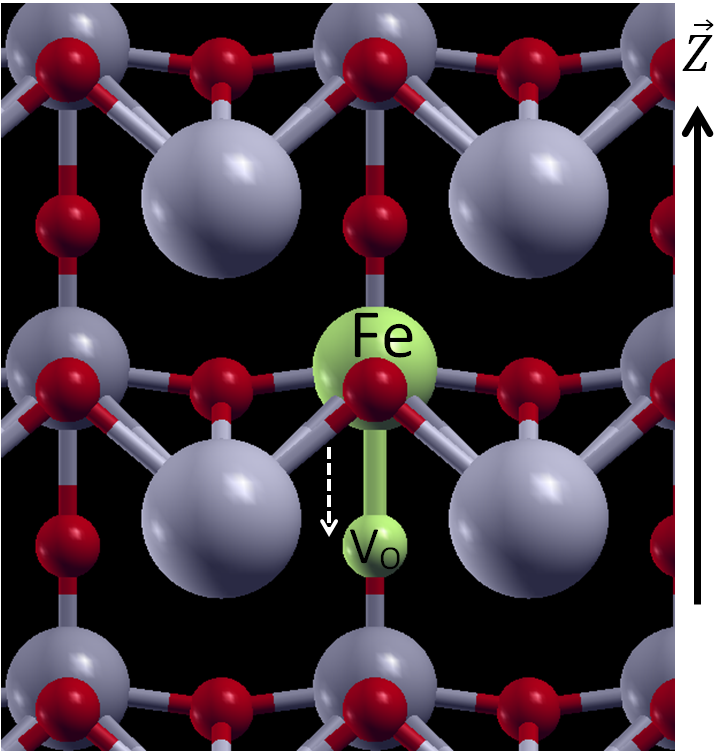}
  \label{fig:3}
}
\caption{The different configurations for the Fe$^{'}_{Ti}$-V$^{^{\textbf{..}}}_O$ defect associates in PbTiO$_3$. The lattice polarization is along the positive z direction, as shown by the black arrow, while the expected defect polarization in each configuration is indicated by the small dashed arrow.}
\label{figFeVO}
 \end{figure*}

\begin{figure}[!ht]
\centering
\includegraphics[width=0.5\textwidth]{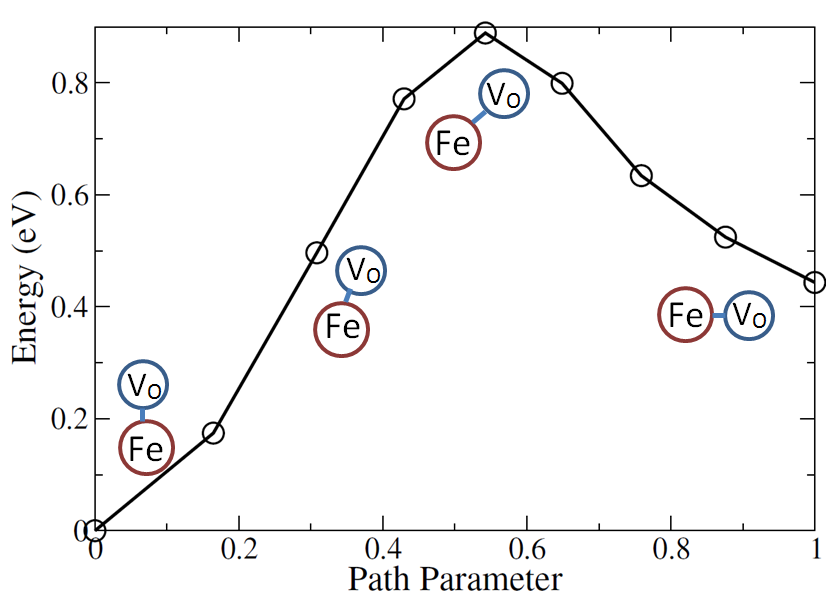}
\caption{Energy barrier for the hopping of an oxygen vacancy from the axial to the equatorial configuration.}
\label{nebFeVO}
 \end{figure}
 
Due to easy formation of lead and oxygen vacancies \cite{Yao2011} in lead-based perovskites it is very likely that the V$^{''}_{Pb}$-V$^{^{\textbf{..}}}_O$ divacancy could exist in an undoped material \cite{Dai1991}.However, the structure of this divacancy is still controversial. Some authors claim the nearest-neighbor configuration to be the most stable\cite{Cockayne2004}, while others claim the two vacancies to be located further apart\cite{Poykko2000}. Cockayne et al.\cite{Cockayne2004} calculated the polarization of the nearest-neighbor divacancy using the Berry phase approach\cite{Resta1994} and found it to be three times the bulk polarization. Here, we study the stability of different configurations of the lead-oxygen divacancy to understand its interactions with domain walls. The supercells were charge neutral since the lead vacancy is doubly (negatively) charged and the oxygen vacancy is doubly (positively) charged. Our calculations reveal that the next-nearest-neighbor axial configuration to be the ground-state, as shown in Fig.\ref{VPbVO3}. The nearest neighbor axial configuration depicted in Fig.\ref{VPbVO1} is the next most stable. The equatorial configuration also shows a tendency to align along the polarization, as shown in Fig.\ref{VPbVO2}, while the equatorial configuration opposing the  polarization depicted in Fig.\ref{VPbVO-opp}, is the least stable. Hence, we see a similarity between the behavior of the V$^{''}_{Pb}$-V$^{^{\textbf{..}}}_O$ divancancy and the Fe$^{'}_{Ti}$-V$^{^{\textbf{..}}}_O$ defect associate, where these defects are oriented preferably in the direction of polarization, explaining why nominally pure material containing lead and oxygen vacancies behave like hard, acceptor-doped materials.

 \begin{figure*}[!ht]
\centering
\subfigure[The next nearest-neighbor axial configuration of the  V$^{''}_{Pb}$-V$^{^{\textbf{..}}}_O$ divacancy is the ground-state structure. In this case there is a component of polarization pointing in the positive z direction which is the direction of the lattice polarization.]
{
 \includegraphics[width=0.315\textwidth]{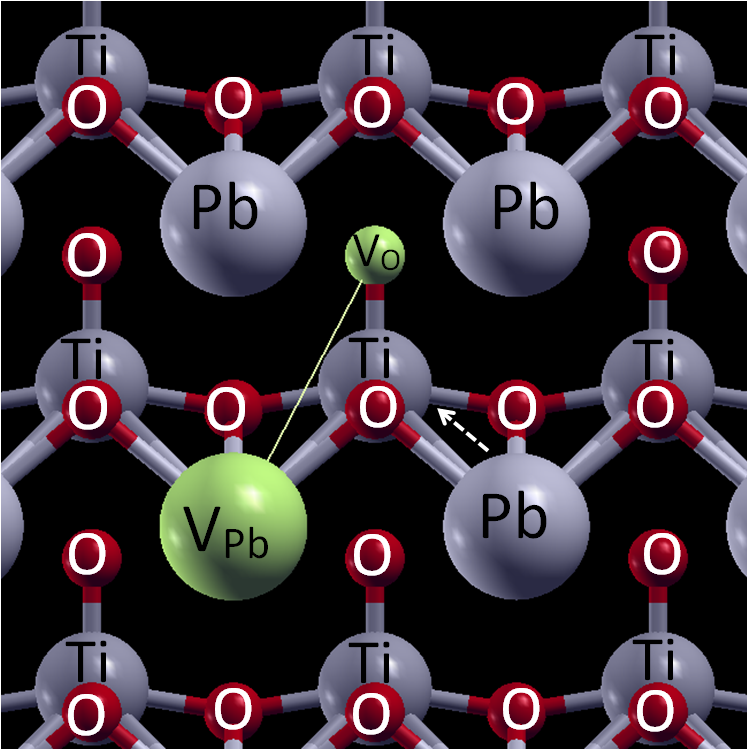}
  \label{VPbVO3}
}
\subfigure[Nearest-neighbor axial configuration = 0.17 eV. In general it was observed that the oxygen vacancy prefers to be on an axial site.]
{
 \includegraphics[width=0.315\textwidth]{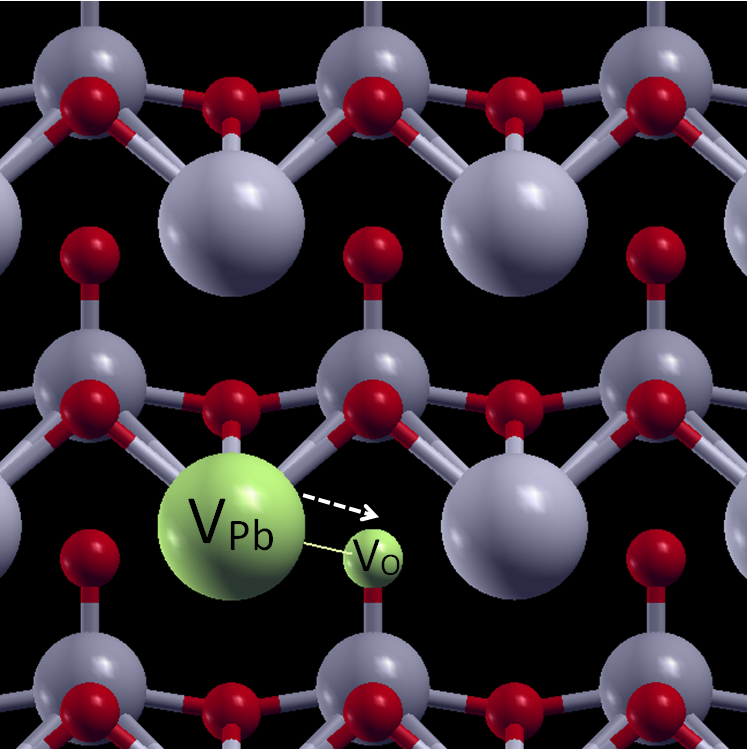}
  \label{VPbVO1}
}
\subfigure[Nearest-neighbor equatorial configuration = 0.30 eV. Even in this case, theres is a component of polarization pointing in the positive z direction.]
{
 \includegraphics[width=0.315\textwidth]{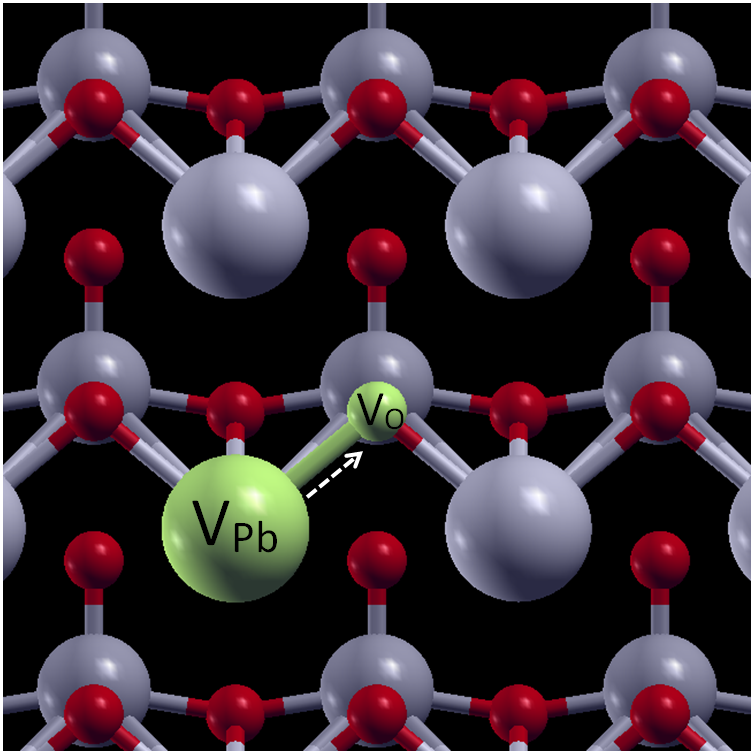}
  \label{VPbVO2}
}
\subfigure[Nearest-neighbor equatorial configuration with opposite polarization = 0.63 eV.]
{
 \includegraphics[width=0.315\textwidth]{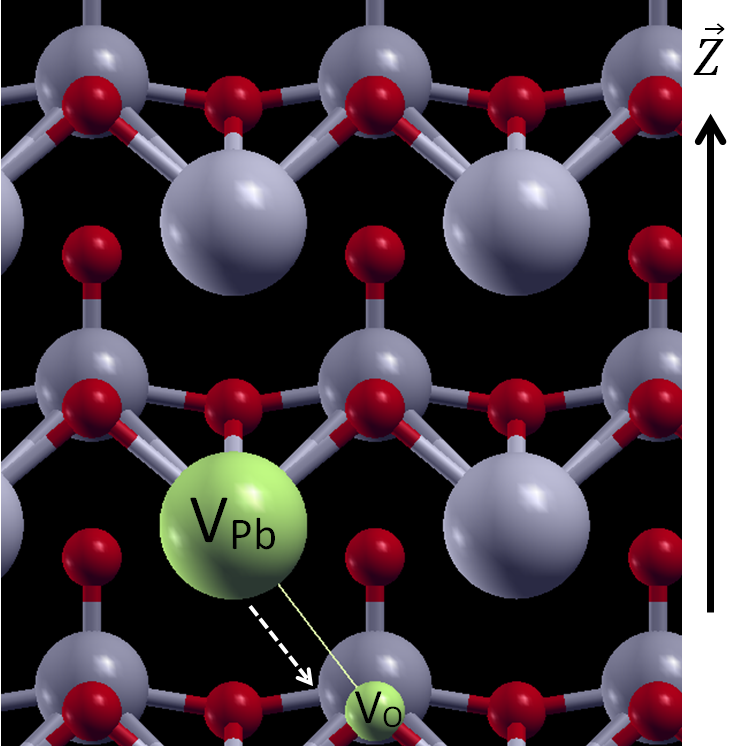}
  \label{VPbVO-opp}
}
\caption{Different configurations of the V$^{''}_{Pb}$-V$^{^{\textbf{..}}}_O$ defect associate relative to the lattice polarization shown with the black arrow. The expected defect polarization in each configuration is indicated by the small dashed arrow. }
\label{figVPbVO}
 \end{figure*}

\section{Donor - Lead vacancy defect associate}

Donor-doped PZT is thought to be charge compensated by the formation of lead vacancies \cite{Gerson1960,Mackie2010,Dai1991}. Currently, this defect complex has not been characterized although many conjectures on softening have been made, based on the formation of a Nb$^{^{\textbf{.}}}_{Ti}$-V$^{''}_{Pb}$ associate. Here we investigate if this defect complex exhibits similar properties to the Fe$^{'}_{Ti}$-V$^{^{\textbf{..}}}_O$ complex investigated earlier. For this system the supercell has a net negative charge, assuming the lead vacancy to be doubly negatively charged and the niobium substitutional defect to have a single positive charge. Fig.\ref{figNbVPb} shows the different configurations of the defect associate along with the relative differences in energy with respect to the ground-state. Fig.\ref{NbVPb1}  depicts a schematic representation of the defect complex oriented along the direction of polarization, with the niobium substitutional defect at the center of the supercell. The configuration which is oriented away from the direction of polarization is shown in Fig.\ref{NbVPb2}, and a configuration in which the two defects are located further away from each other is presented in Fig.\ref{NbVPbiso}. From these results we find that the selectivity for different configurations is much less than in acceptor-doped materials. Partial alignment for the defect associate in the direction of polarization is only 0.06 eV lower in energy than when oriented away from the bulk polarization. Even more interestingly, we find that that this defect complex is not tightly bound, unlike the acceptor defect associate, with a formation energy of just 0.04 eV (Fig.\ref{NbVPbiso}). EPR studies on Gd donor-doped soft PZT also seem to show no coupling with lead vacancies\cite{Eichel2006}, and our results on PbTiO$_3$ agree well with this observation. 

 \begin{figure*}[!ht]
\centering
\subfigure[The ground-state configuration of the Nb$^{^{\textbf{.}}}_{Ti}$-V$^{''}_{Pb}$ defect associate oriented in the direction of lattice polarization.]
{
 \includegraphics[width=0.315\textwidth]{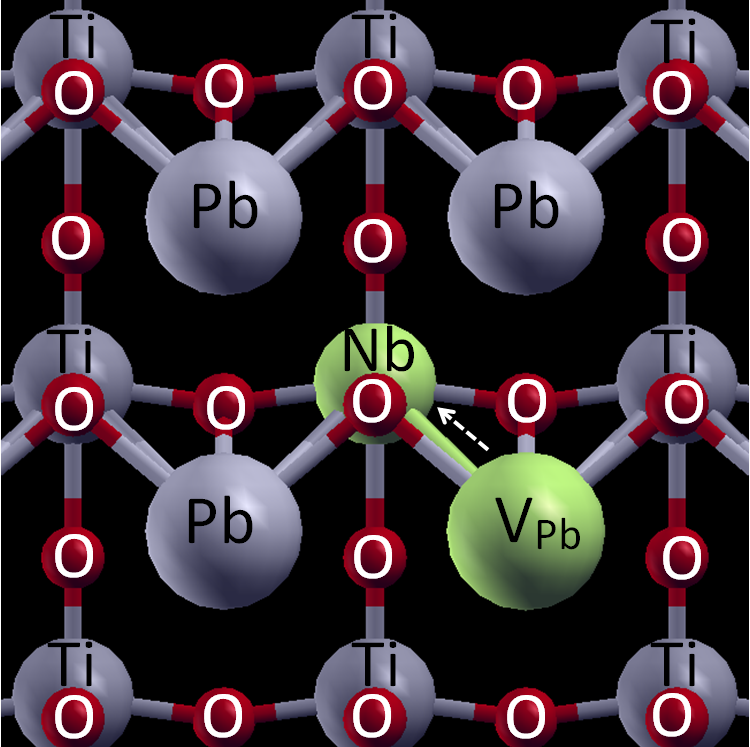}
  \label{NbVPb1}
}
\subfigure[In this case the Nb$^{^{\textbf{.}}}_{Ti}$-V$^{''}_{Pb}$ defect associate is oriented away from the direction of lattice polarization. It has an energy of just from polarization 0.06 eV higher than the ground-state.]
{
 \includegraphics[width=0.315\textwidth]{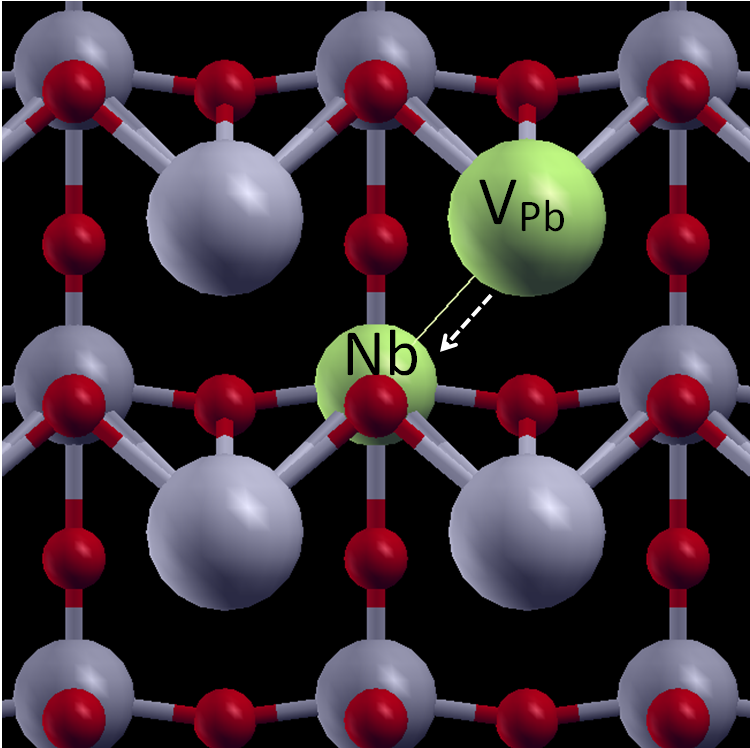}
 \label{NbVPb2}
}
\subfigure[Dissociated defects = 0.04 eV higher than ground-state.]
{
 \includegraphics[width=0.315\textwidth]{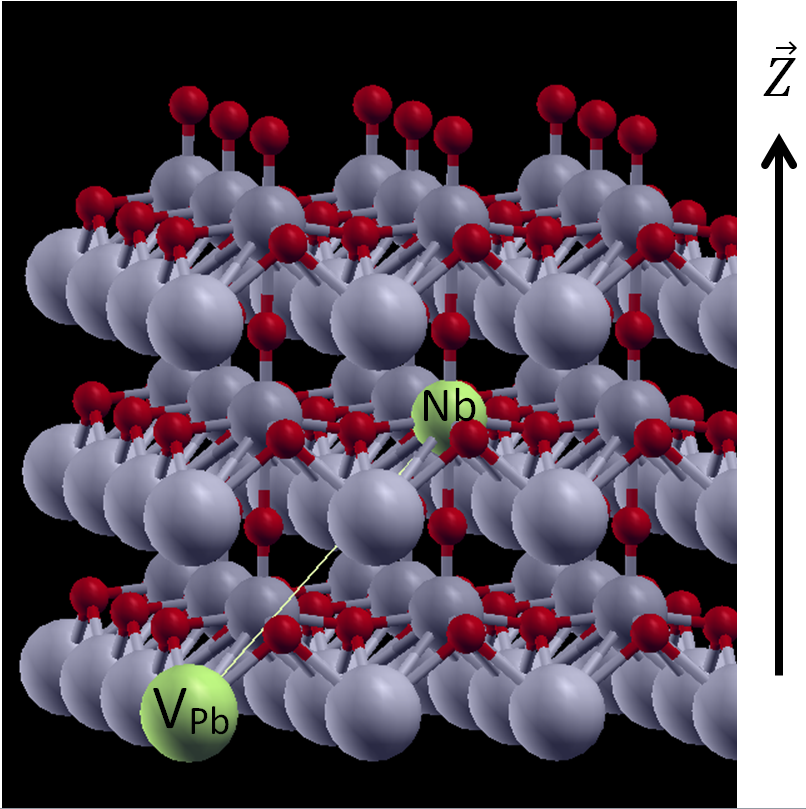}
 \label{NbVPbiso}
}
\caption{Different configurations of Nb$^{^{\textbf{.}}}_{Ti}$-V$^{''}_{Pb}$. There is almost no difference in the energy between the configuration oriented in the direction of lattice polarization, (a), and the one oriented away from the polarization (b). Furthermore, this defect associate does not have significant binding energy as shown in (c).}
\label{figNbVPb}
 \end{figure*}

\section{Defect-domain wall interaction}
In order to investigate the effect of dopants on domain walls, the structure and formation energy of a 180$^o$ domain wall in pure PbTiO$_3$ was calculated. Fig.\ref{180dw} shows the schematics for this domain wall, centered on the (100) plane of lead and oxygen atoms. The formation energy is 116 mJ/m$^2$ and the barrier energy for the movement of the domain wall from one plane to the next is 28 mJ/m$^2$. The titanium centered configuration is the saddle point as shown in Fig.\ref{neb24cell-label}: The energies are plotted in eV and since we know the area of the domain wall in the supercell we report the barrier energy in mJ/m$^2$. These barrier energies reported here are slightly lower than that obtained by Meyer et al. \cite{Meyer2002}, because in the present case the nudged elastic band is used to find the minimum energy pathway for domain wall movement, rather than using fixed atomic position calculations. This 180$^o$ domain wall is extremely sharp, extending only one unit cell in either direction of the interface before recovering bulk positions.

To see how this barrier energy is affected by the presence of ordered defects, we repeated the nudged elastic band calculations with ordered defects at the domain wall. The supercell was similar to that shown in Fig.\ref{180dw} but the dimensions in the y and z direction are doubled, with supercells made by 6*2*2 unit cells.
\begin{figure}[!ht]
\centering
\includegraphics[width=0.5\textwidth]{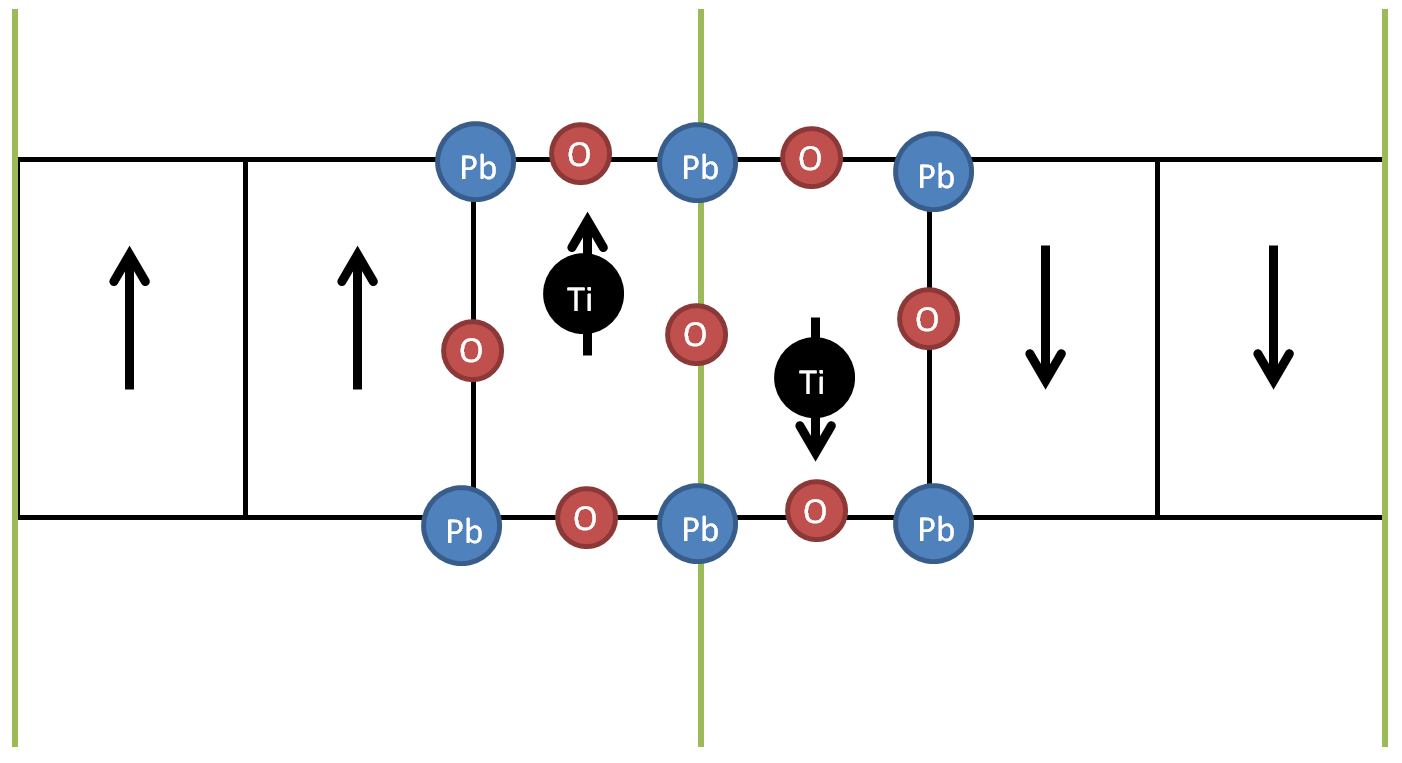}
\caption{Schematics of a 180$^o$ domain wall supercell with arrows showing the direction of lattice polarization in each cell.}
\label{180dw}
 \end{figure}
 
 \begin{figure}[!ht]
\centering
\includegraphics[width=0.5\textwidth]{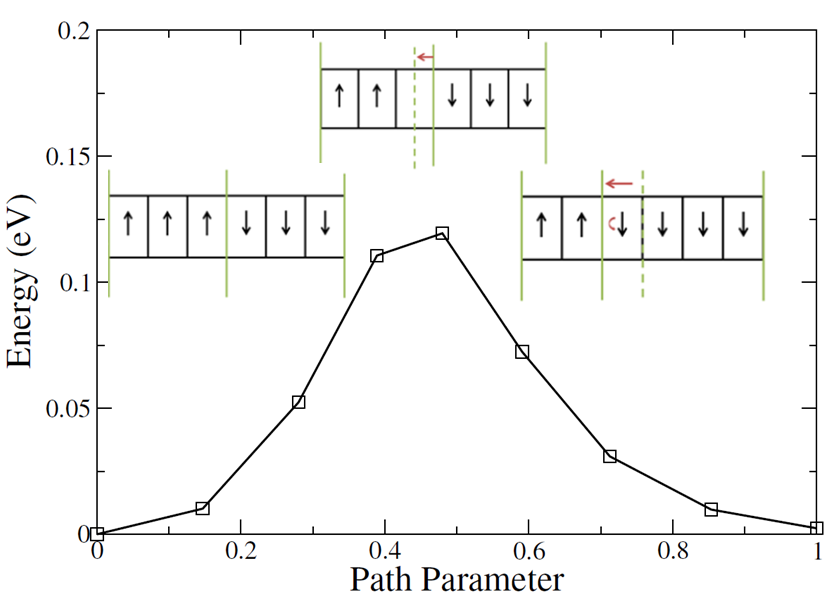}
\caption{Barrier energy for the movement of the domain wall across one unit cell calculated using the nudged elastic band method. The three domain configurations (from left to right) correspond to starting,intermediate and final positions respectively.}
\label{neb24cell-label}
 \end{figure}

With this set up, we study the pinning effect of the Fe$^{'}_{Ti}$-V$^{^{\textbf{..}}}_O$ defect complex  and the V$^{''}_{Pb}$-V$^{^{\textbf{..}}}_O$ divacancy. The Fe$^{'}_{Ti}$-V$^{^{\textbf{..}}}_O$ defect associate is oriented in the +z direction (as depicted in Fig. \ref{fig:1}) and the V$^{''}_{Pb}$-V$^{^{\textbf{..}}}_O$ defect associate also has a component in this direction (as shown in Fig.\ref{VPbVO2}. 

Fig.\ref{schFeVO} shows a schematic representation of the movement of the domain wall across the unit cell containing the defect associate. Once the domain wall moves, the defect polarization is oriented in the opposite direction of the lattice polarization. The nudged elastic band method was used to calculate the minimum energy pathway for this process. Fig.\ref{orddef} depicts the barrier energy in the presence of defects compared to the pure undoped case.

 \begin{figure}[!ht]
\centering
\subfigure[]
{
 \includegraphics[width=0.21\textwidth]{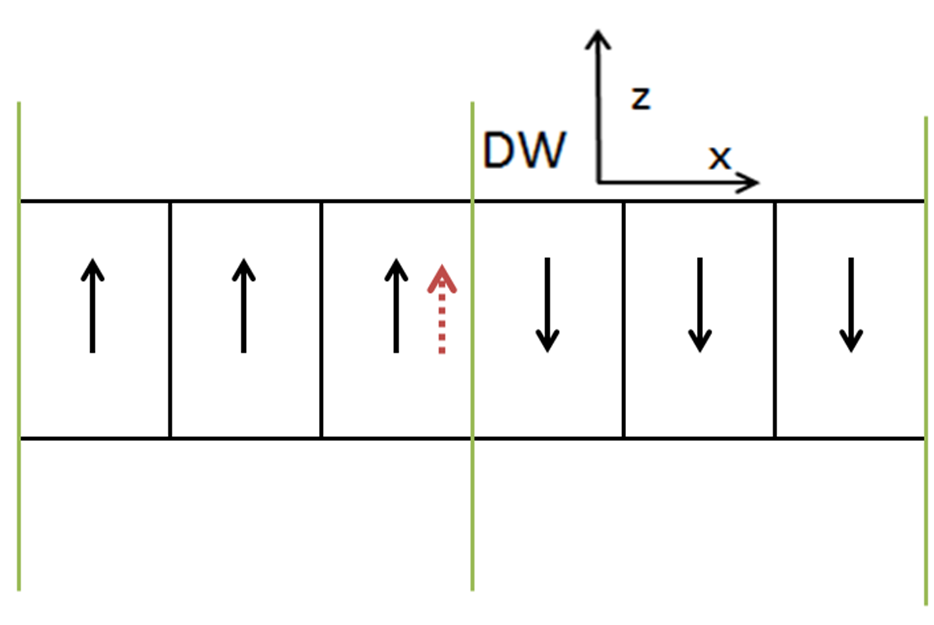}
  \label{sch1FeVO}
}
\subfigure[]
{
 \includegraphics[width=0.21\textwidth]{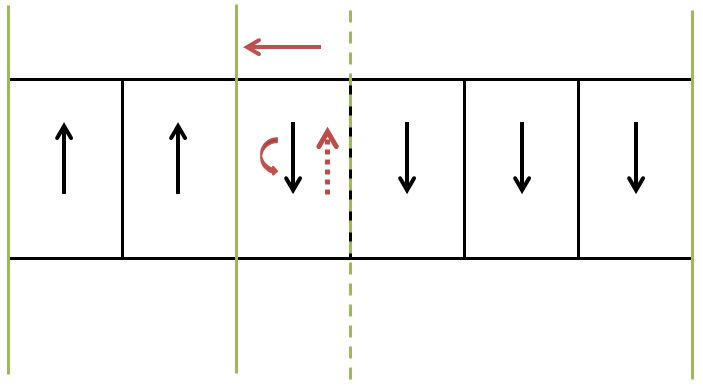}
 \label{sch2FeVO}
}
\caption{Schematic of the movement of the domain wall in the presence of the defect, whose position and polarization are represented by the small dashed arrow. (a)Initial state of supercell with defect oriented in the direction of the cell polarization. (b)Final state after domain wall motion where the defect is now oriented in the opposite direction to cell polarization.}
\label{schFeVO}
 \end{figure}

\begin{figure}[!ht]
\centering
\includegraphics[width=0.4\textwidth]{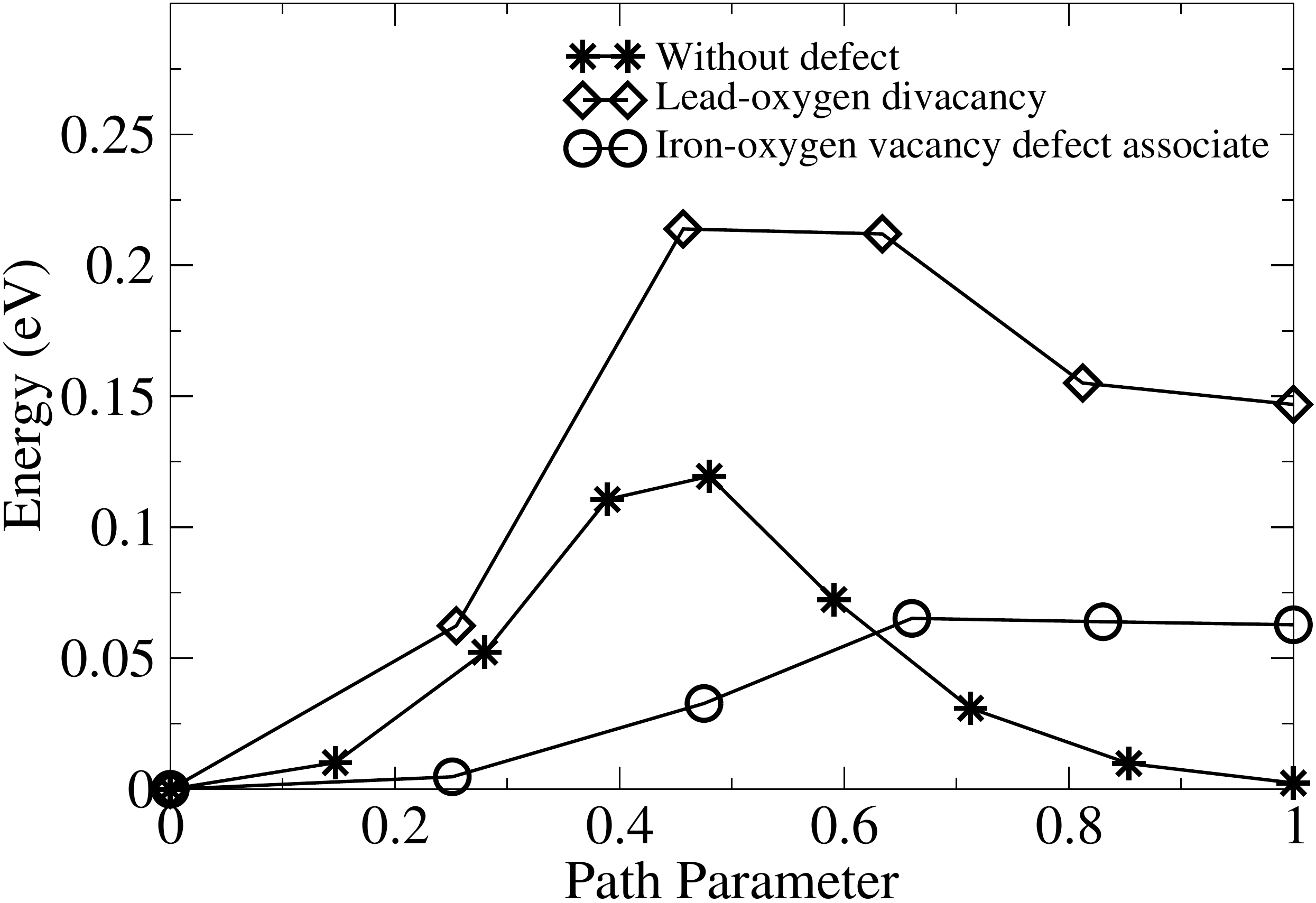}
\caption{Barrier energy of domain wall movement in the presence of defects.}
\label{orddef}
 \end{figure}

It is clear that there is an asymmetry in the barrier energy in both defect cases because of the change in the relative polarization of the defect associates and the bulk. The V$^{''}_{Pb}$-V$^{^{\textbf{..}}}_O$ divacancy also has a strong effect on the height of the barrier for the movement of the domain wall. To understand better the energy profile across the entire supercell, further calculations were performed to find out the position of these defects relative to the domain wall. Fig.\ref{FeVOdomint} plots the energy of the supercell for four different positions of the Fe$^{'}_{Ti}$-V$^{^{\textbf{..}}}_O$  defect associate with respect to the domain wall. The numbers on the x axis represent different configurations which vary in relative positions to this domain wall as shown in Fig.\ref{legendpos}. \emph{Position -1} and \emph{Position 0} represent configurations in which the defect polarization and lattice polarization are in the same direction. \emph{Position 1} and \emph{Position 2} are configurations in which the two polarizations are in opposite directions. Both \emph{Position 0} and \emph{Position 1} are at the interface of the domain wall. There are two main observations from these calculations. First, the defect associate is least stable at the interior of the domain with opposing polarization. This indeed points to a strong internal field effect which forces the domain wall to move. Second, the defect associate is more stable at the interface of the domain wall rather than the interior of the domain with the same polarization direction. Hence it seems pinning is a combination of both bulk and domain wall effects.

The same calculations were repeated for the V$^{''}_{Pb}$-V$^{^{\textbf{..}}}_O$ defect associate and a similar profile was obtained (Fig.\ref{vpbvodom}). The reduction in energy at \emph{Position 2} arises since the pinning force is so high that the domain wall will shift its position relative to the defect. This result also agrees with strong domain wall pinning observed in undoped PZT. To complete the study we also calculated the relative stability of isolated defects relative to the domain wall. In general, it was observed that all defects prefer to be at 180$^o$ domain walls.  He and Vanderbilt\cite{He2003} argued initially that neutral oxygen vacancies are more stable at such domain walls, but it has since been shown that oxygen vacancies with a double positive charge are the most stable\cite{Yao2011}. Neutral oxygen vacancies have a formation energy of around 10 eV\cite{He2003} but doubly charged oxygen vacancies have been reported with a formation energy of just 0.28 eV\cite{Yao2011} or even as low as -3.76 eV\cite{Shimada2013}. Hence, doubly positive-charged oxygen vacancies are investigated in this study. 
Fig.\ref{VOpos} shows the relative stability of the various oxygen vacancies (i.e x-V$_O$, y-V$_O$ and z-V$_O$) at different positions from the domain wall. The energies are plotted with respect to the ground state of an oxygen vacancy in the z direction at the domain wall (z-V$_O$). From this potential energy surface we see that all types of oxygen vacancies have a lower energy at the domain wall. Fig.\ref{Defpos} shows a similar plot for a lead vacancy, a niobium defect on the titanium site, as well as a lanthanum defect on the lead site. All these isolated defects also have a lower energy at the domain wall. The energy difference between \emph{Position 0} (Fig.\ref{pos0}) and \emph{Position 2} (Fig.\ref{pos2}) is taken as a rough estimate of the pinning energy of a defect and Fig.\ref{relpinning} summarizes the pinning energies of defects and defect associates. Considering just the isolated defects, it can be seen that the oxygen vacancies (particularly z-V$_O$ and x-V$_O$)  have the greatest attraction to the domain wall and are hence the strongest pinning centers. However, the oxygen vacancy pinning is three times smaller than that of the Fe$^{'}_{Ti}$-V$^{^{\textbf{..}}}_O$ defect associate, as shown in Fig.\ref{relpinning}. The calculations also indicate that the V$^{''}_{Pb}$-V$^{^{\textbf{..}}}_O$ divacancy could be an even stronger pinning center. As we mentioned earlier, we could not calculate the pinning energy for this defect associate because of the movement of the domain wall to a new equilibrium position in the presence of the V$^{''}_{Pb}$-V$^{^{\textbf{..}}}_O$ defect associate.

 \begin{figure}[H]
\includegraphics[width=0.4\textwidth]{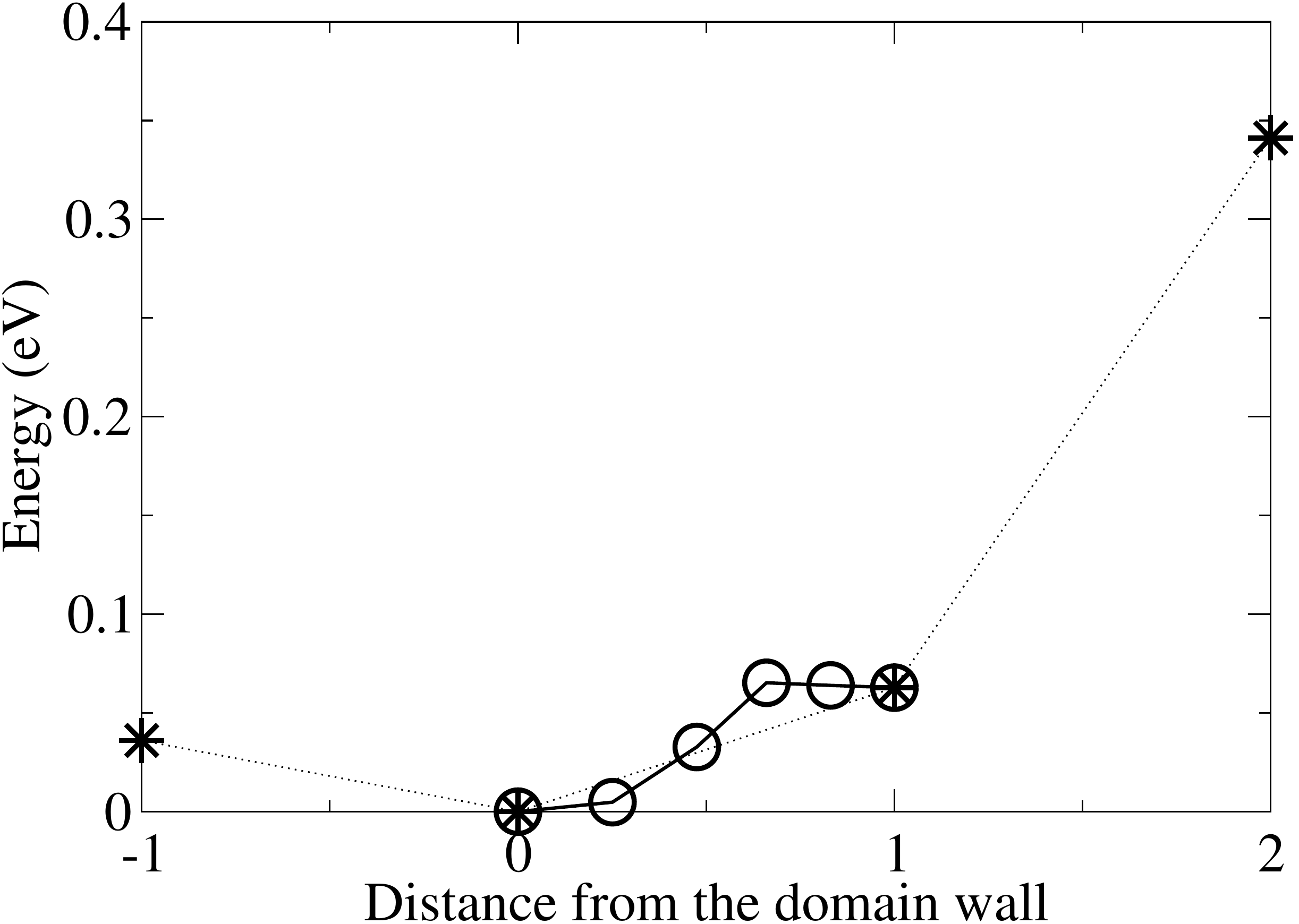}
\caption{Energy profile of the supercell for different positions of Fe$^{'}_{Ti}$-V$^{^{\textbf{..}}}_O$ relative to the domain wall. The labels -1,0,1 and 2 are explained in Fig.\ref{legendpos}. The solid curve shows the NEB calculation for the local movement between Positions 0 and 1. The dashed lines are only guidelines for the eyes.}
\label{FeVOdomint}
 \end{figure}

  \begin{figure}[H]
\centering
\subfigure[ Position -1]
{
 \includegraphics[width=0.22\textwidth]{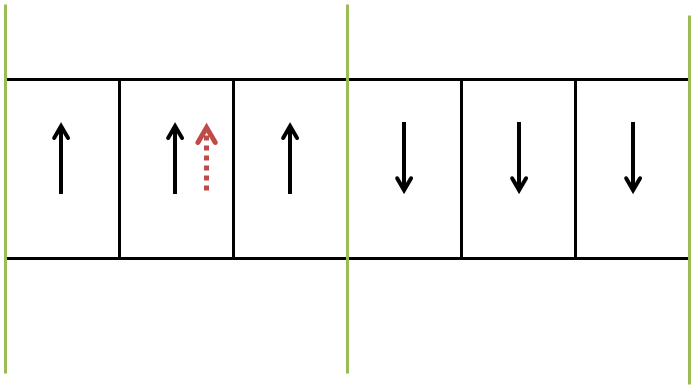}
  \label{pos-1}
}
\subfigure[ Position 0 ]
{
 \includegraphics[width=0.22\textwidth]{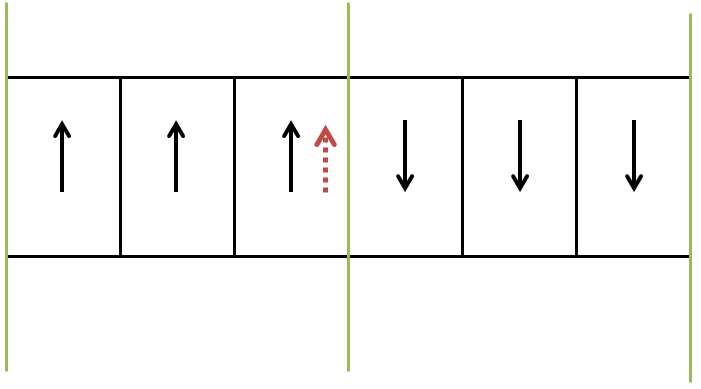}
 \label{pos0}
}
\subfigure[ Position 1 ]
{
 \includegraphics[width=0.22\textwidth]{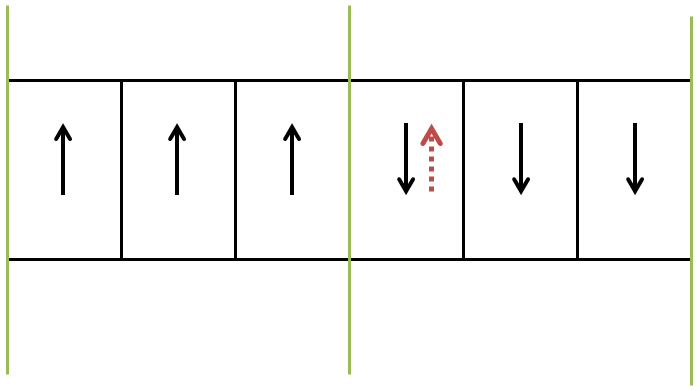}
 \label{pos1}
}
\subfigure[ Position 2 ]
{
 \includegraphics[width=0.22\textwidth]{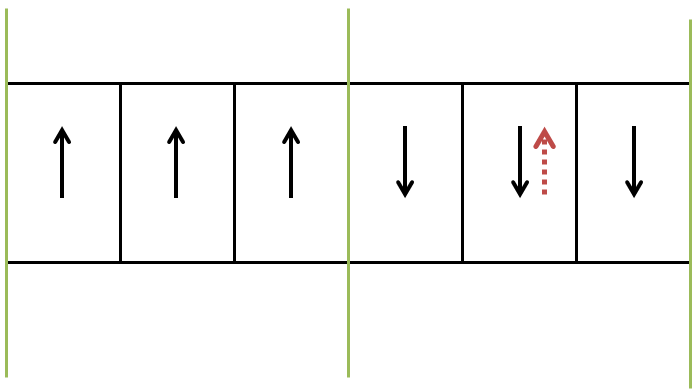}
 \label{pos2}
}
\caption{Configurations which differ in relative distance between defect and domain wall. The defect position and polarization is represented by the small dashed arrow.}
\label{legendpos}
 \end{figure}

 \begin{figure}[!ht]

\includegraphics[width=0.4\textwidth]{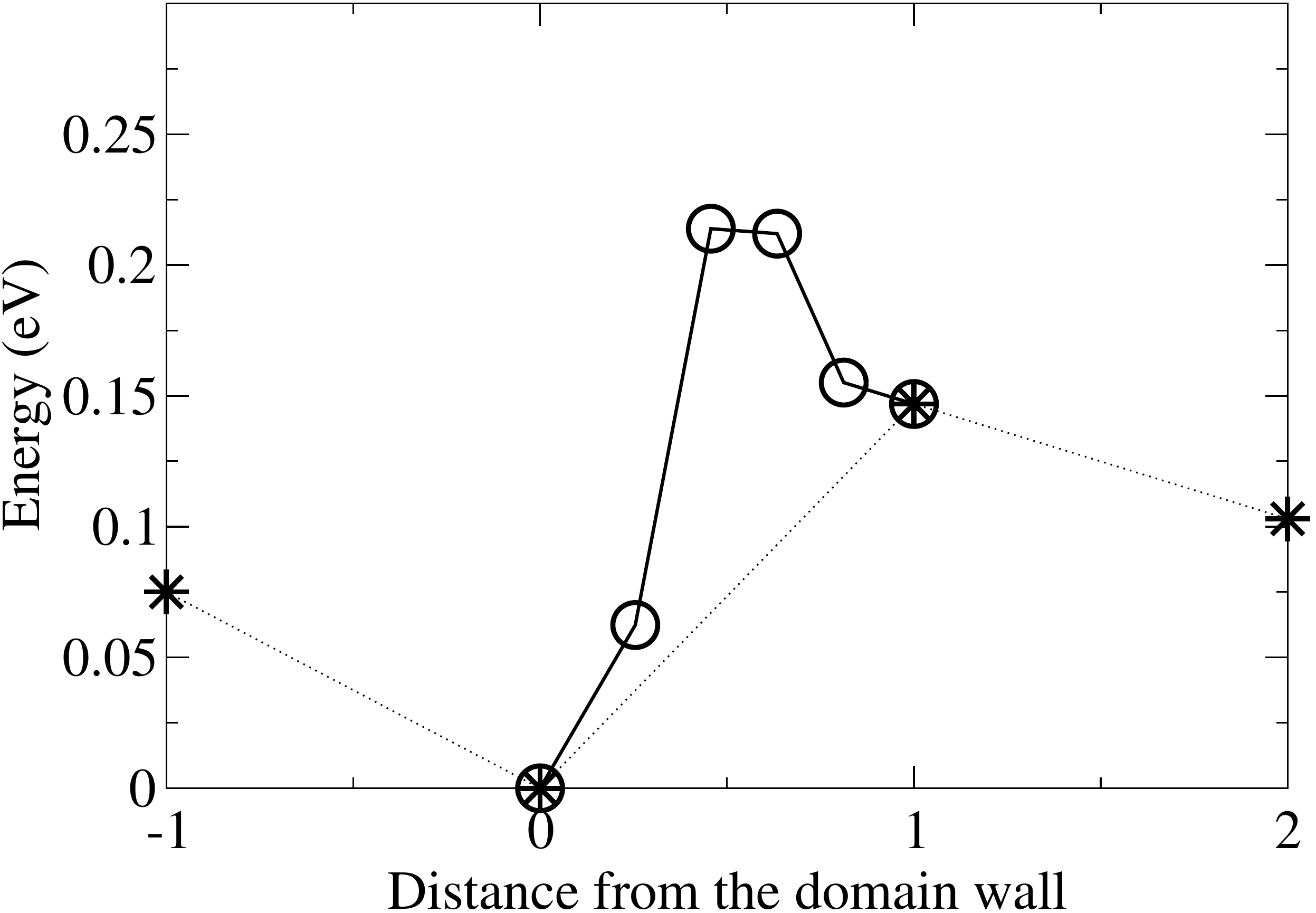}
\caption{Energy profile of the supercell for different positions of V$^{''}_{Pb}$-V$^{^{\textbf{..}}}_O$ relative to the domain wall.  The solid curve shows the NEB calculation for the local movement of domain wall between Positions 0 and 1. The reduction in energy for \emph{Position 2} is because the pinning strength is so high that it forces the domain wall to move towards the defect.}
\label{vpbvodom}
 \end{figure}
 
  \begin{figure}[!ht]

\includegraphics[width=0.4\textwidth]{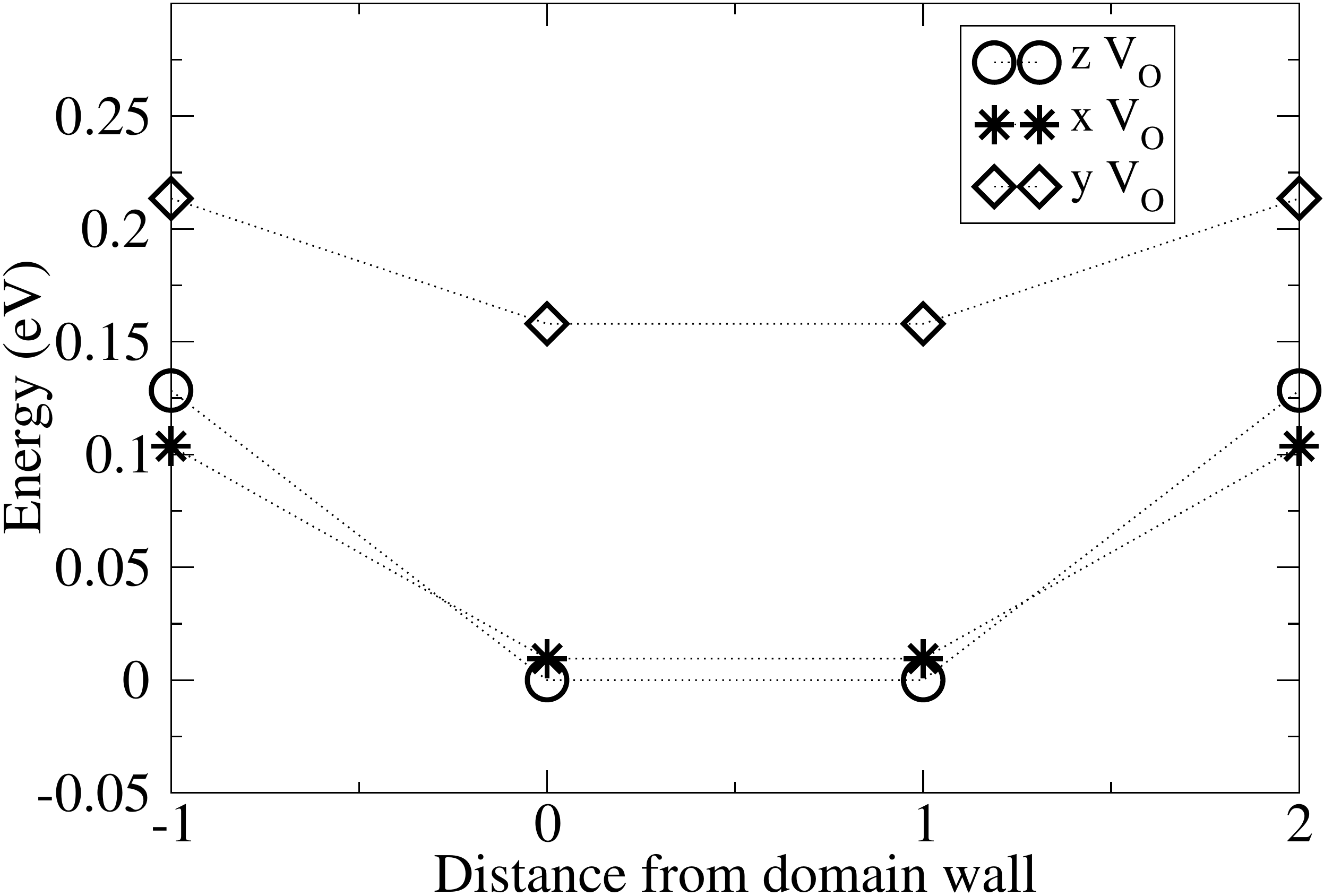}
\caption{Stability of oxygen vacancies at different distances from the domain wall.}
\label{VOpos}
 \end{figure}
 
  \begin{figure}[!ht]

\includegraphics[width=0.4\textwidth]{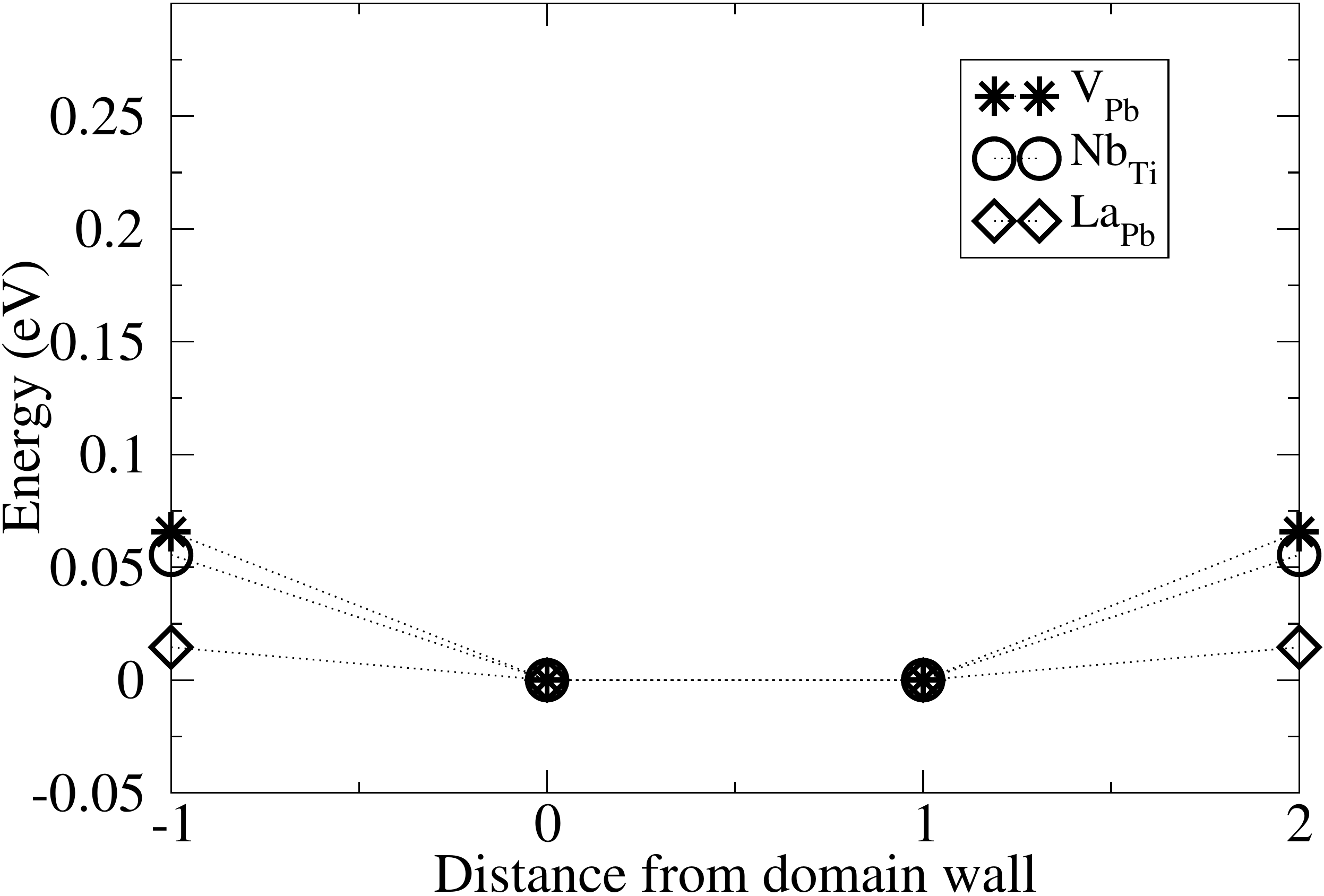}
\caption{Stability of lead vacancy, substitutional niobium defect and substitutional lanthanum defect at different distances from the domain wall. The energies are plotted with respect to the ground-state structure in each case.}
\label{Defpos}
 \end{figure}
 
 \begin{figure}[!ht]
 \includegraphics[width=0.4\textwidth]{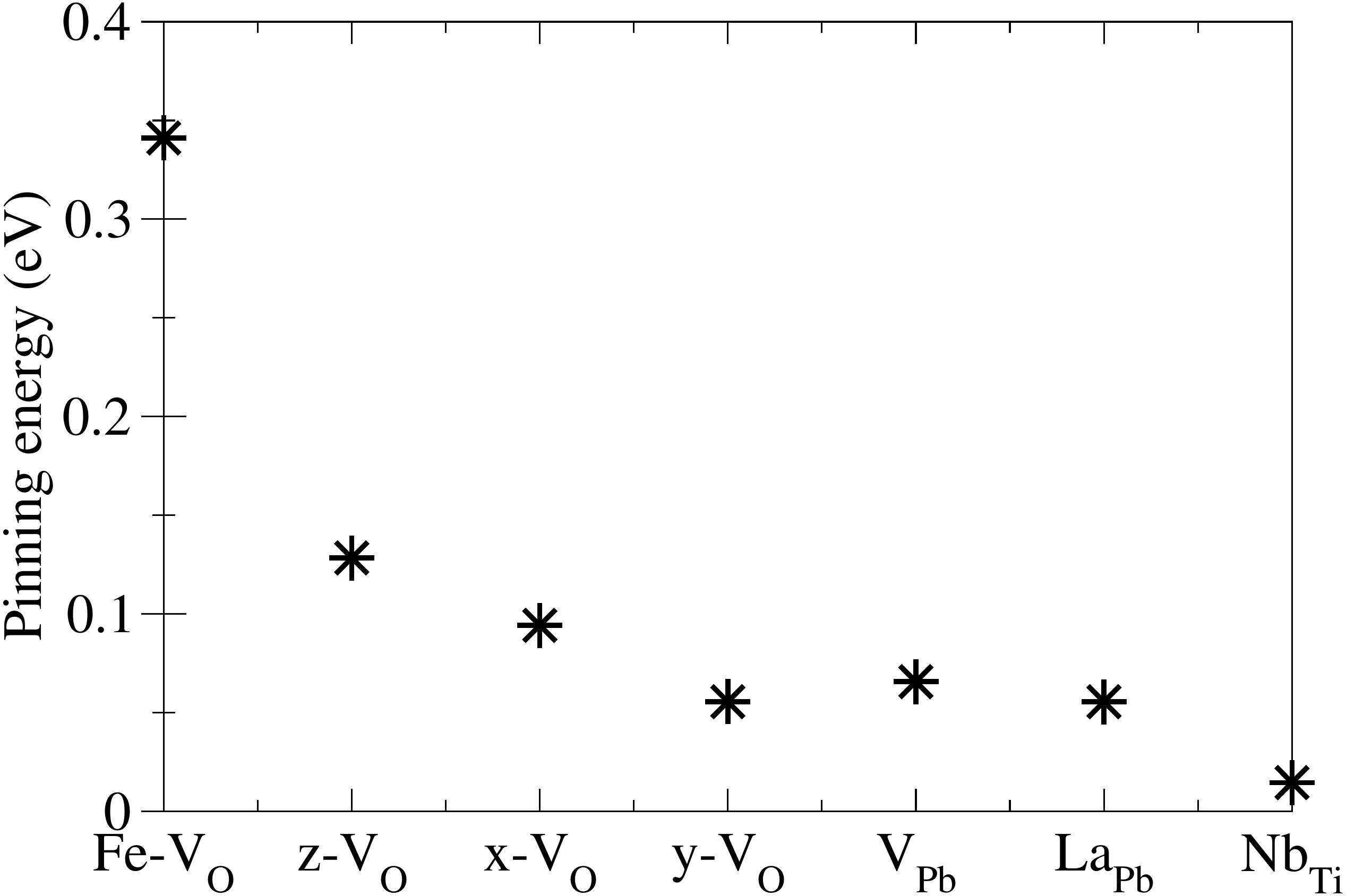}
\caption{Pinning energies (see text for definition) of defects and defect associates.}
\label{relpinning}
 \end{figure}

\section{Conclusions}

 In Fe acceptor-doped PbTiO$_3$, a defect associate is formed between the iron substitutional defect and a charged oxygen vacancy (Fe$^{'}_{Ti}$-V$^{^{\textbf{..}}}_O$). This defect associate aligns in the direction of the bulk polarization. This alignment is due to both electrostatic and elastic effects. The polarization direction of the defect associate can be changed by the hopping of oxygen vacancies and nudged elastic band calculations reveal that the activation energy for this process matches that of the experimentally determined AC conductivity and deaging process in hard PZT. A similar ordering phenomena is also observed in the case of the lead-oxygen divacancy. However, in the case of Nb donor-doped PbTiO$_3$, the defect associate between the niobium substitutional defect and lead vacancy shows no binding energy and no preferential alignment with the polarization. Hence, the Nb$^{^{\textbf{.}}}_{Ti}$-V$^{''}_{Pb}$ defect associate is unlikely to exist and even if such defect complexes exist, it is clear that they do not interact strongly with the lattice polarization. It was also observed that ordered defect associates have a very strong effect on 180$^o$ domain walls. They are not only more stable at the domain wall but they also exhibit the characteristics of a random field defect, i.e. they break the degeneracy of polarization states to prefer a certain orientation. Oxygen vacancies, lead vacancies, niobium substitutional defects, and lanthanum substitutional defects are also more stable at the domain wall. Amongst isolated defects, oxygen vacancies showed the greatest preference to be at the domain wall. However, defect associates showed 3 times higher pinning strength compared to lone oxygen vacancies. 
 
 Finally, it is clear from this work that oxygen vacancies are key in forming polar defect complexes leading to pinched hysteresis loops and ageing in both undoped and hard PZT. It is also shown that both the ``bulk effect'' and ``domain-wall effect'' are likely to contribute to the hardening phenomenon. The fact that polar defect associates are more stable at 180$^o$ domain walls has interesting consequences. It may explain why nano-domains are observed in Fe doped bulk ceramics\cite{Jin2009}. The possibility of tuning domain size and domain orientation by the orientation of defect dipoles is technologically intriguing: Controlled nanodomains could produce higher density FeRAM storage devices. It may also give rise to the possibility of stabilizing charged domain walls which would otherwise be unstable. Recently, it has been shown that it is possible to have metallic conductivity at charged nano-domain walls in PZT thin films\cite{Maksymovych2012} and this is something which would be very interesting to characterize from first-principles.
 
 A better understanding on the issue of softening has also been obtained. Due the basic principle of electro-neutrality, donor dopants are expected to reduce the concentration of oxygen vacancies; there have also been some first-principles calculation showing an increased formation energy of oxygen vacancies in the presence of donor dopants\cite{Zhang2006a}. From these observations we point out that it is likely that donor-doped samples have increased domain mobilities due to a lower concentration of oxygen vacancies, leading to an absence of polar defect complexes. However, this may not be the only mechanism of softening. The effect of these donor dopants on 90$^o$ domain walls (of great interest for piezoelectric applications) is yet to be investigated, and would be required in the future for a complete study. 
 
\section{Acknowledgements}
A.C. and N.S. acknowledge support received from the European Research Council under the EU 7th Framework Programme (FP7/2007-2013)/ ERC grant agreement n\textsuperscript{o}[268058].

\clearpage
%\begin{acknowledgments}
%\end{acknowledgments}
\bibliographystyle{apsrev4-1}
\bibliography{ref}

%Merlin.mbs v4.21 2009-07-09.
\providecommand{\noopsort}[1]{}\providecommand{\singleletter}[1]{#1}%
\begin{thebibliography}{10}%
\makeatletter
\providecommand \@ifxundefined [1]{%
 \ifx #1\undefined \expandafter \@firstoftwo
 \else \expandafter \@secondoftwo
\fi
}%
\providecommand \@ifnum [1]{%
 \ifnum #1\expandafter \@firstoftwo
 \else \expandafter \@secondoftwo
\fi
}%
\providecommand \enquote [1]{``#1''}%
\providecommand \bibnamefont  [1]{#1}%
\providecommand \bibfnamefont [1]{#1}%
\providecommand \citenamefont [1]{#1}%
\providecommand\href[0]{\@sanitize\@href}%
\providecommand\@href[1]{\endgroup\@@startlink{#1}\endgroup\@@href}%
\providecommand\@@href[1]{#1\@@endlink}%
\providecommand \@sanitize [0]{\begingroup\catcode`\&12\catcode`\#12\relax}%
\@ifxundefined \pdfoutput {\@firstoftwo}{%
 \@ifnum{\z@=\pdfoutput}{\@firstoftwo}{\@secondoftwo}%
}{%
 \providecommand\@@startlink[1]{\leavevmode\special{html:<a href="#1">}}%
 \providecommand\@@endlink[0]{\special{html:</a>}}%
}{%
 \providecommand\@@startlink[1]{%
  \leavevmode
  \pdfstartlink
   attr{/Border[0 0 1 ]/H/I/C[0 1 1]}%
   user{/Subtype/Link/A<</Type/Action/S/URI/URI(#1)>>}%
  \relax
 }%
 \providecommand\@@endlink[0]{\pdfendlink}%
}%
\providecommand \url  [0]{\begingroup\@sanitize \@url }%
\providecommand \@url [1]{\endgroup\@href {#1}{\urlprefix}}%
\providecommand \urlprefix [0]{URL }%
\providecommand \Eprint[0]{\href }%
\@ifxundefined \urlstyle {%
  \providecommand \doi [1]{doi:\discretionary{}{}{}#1}%
}{%
  \providecommand \doi [0]{doi:\discretionary{}{}{}\begingroup
  \urlstyle{rm}\Url }%
}%
\providecommand \doibase [0]{http://dx.doi.org/}%
\providecommand \Doi[1]{\href{\doibase#1}}%
\providecommand \bibAnnote [3]{%
  \BibitemShut{#1}%
  \begin{quotation}\noindent
    \textsc{Key:}\ #2\\\textsc{Annotation:}\ #3%
  \end{quotation}%
}%
\providecommand \bibAnnoteFile [2]{%
  \IfFileExists{#2}{\bibAnnote {#1} {#2} {\input{#2}}}{}%
}%
\providecommand \typeout [0]{\immediate \write \m@ne }%
\providecommand \selectlanguage [0]{\@gobble}%
\providecommand \bibinfo [0]{\@secondoftwo}%
\providecommand \bibfield [0]{\@secondoftwo}%
\providecommand \translation [1]{[#1]}%
\providecommand \BibitemOpen[0]{}%
\providecommand \bibitemStop [0]{}%
\providecommand \bibitemNoStop [0]{.\EOS\space}%
\providecommand \EOS [0]{\spacefactor3000\relax}%
\providecommand \BibitemShut [1]{\csname bibitem#1\endcsname}%
%</preamble>
\bibitem{Jaffe1971}%
  \BibitemOpen
  \bibfield{author}{%
  \bibinfo {author} {\bibfnamefont{B.}~\bibnamefont{Jaffe}}, \bibinfo {author}
  {\bibfnamefont{W.~R.}\ \bibnamefont{Cook}},\ and\ \bibinfo {author}
  {\bibfnamefont{H.}~\bibnamefont{Jaffe}},\ }%
  \emph{\bibinfo {title} {Piezoelectric ceramics}}\ (\bibinfo {publisher}
  {Academic press London},\ \bibinfo {year} {1971})%
  \bibAnnoteFile{NoStop}{Jaffe1971}%
\bibitem{Lambeck1986}%
  \BibitemOpen
  \bibfield{author}{%
  \bibinfo {author} {\bibfnamefont{P.}~\bibnamefont{Lambeck}}\ and\ \bibinfo
  {author} {\bibfnamefont{G.}~\bibnamefont{Jonker}},\ }%
  \bibfield{journal}{%
  \Doi{10.1016/0022-3697(86)90042-9}{\bibinfo {journal} {Journal of Physics and
  Chemistry of Solids}}\ }%
  \textbf{\bibinfo {volume} {47}},\ \bibinfo {pages} {453 } (\bibinfo {year}
  {1986}),\ ISSN \bibinfo {issn} {0022-3697}%
  \bibAnnoteFile{NoStop}{Lambeck1986}%
\bibitem{Zhang2005}%
  \BibitemOpen
  \bibfield{author}{%
  \bibinfo {author} {\bibfnamefont{L.~X.}\ \bibnamefont{Zhang}}\ and\ \bibinfo
  {author} {\bibfnamefont{X.}~\bibnamefont{Ren}},\ }%
  \bibfield{journal}{%
  \Doi{10.1103/PhysRevB.71.174108}{\bibinfo {journal} {Phys. Rev. B}}\ }%
  \textbf{\bibinfo {volume} {71}},\ \bibinfo {pages} {174108} (\bibinfo {month}
  {May}\ \bibinfo {year} {2005})%
  \bibAnnoteFile{NoStop}{Zhang2005}%
\bibitem{Genenko2007}%
  \BibitemOpen
  \bibfield{author}{%
  \bibinfo {author} {\bibfnamefont{Y.~A.}\ \bibnamefont{Genenko}}\ and\
  \bibinfo {author} {\bibfnamefont{D.~C.}\ \bibnamefont{Lupascu}},\ }%
  \bibfield{journal}{%
  \Doi{10.1103/PhysRevB.75.184107}{\bibinfo {journal} {Phys. Rev. B}}\ }%
  \textbf{\bibinfo {volume} {75}},\ \bibinfo {pages} {184107} (\bibinfo {month}
  {May}\ \bibinfo {year} {2007})%
  \bibAnnoteFile{NoStop}{Genenko2007}%
\bibitem{Genenko2008}%
  \BibitemOpen
  \bibfield{author}{%
  \bibinfo {author} {\bibfnamefont{Y.~A.}\ \bibnamefont{Genenko}},\ }%
  \bibfield{journal}{%
  \Doi{10.1103/PhysRevB.78.214103}{\bibinfo {journal} {Phys. Rev. B}}\ }%
  \textbf{\bibinfo {volume} {78}},\ \bibinfo {pages} {214103} (\bibinfo {month}
  {Dec}\ \bibinfo {year} {2008})%
  \bibAnnoteFile{NoStop}{Genenko2008}%
\bibitem{Feng2008}%
  \BibitemOpen
  \bibfield{author}{%
  \bibinfo {author} {\bibfnamefont{Z.}~\bibnamefont{Feng}}\ and\ \bibinfo
  {author} {\bibfnamefont{X.}~\bibnamefont{Ren}},\ }%
  \bibfield{journal}{%
  \Doi{10.1103/PhysRevB.77.134115}{\bibinfo {journal} {Phys. Rev. B}}\ }%
  \textbf{\bibinfo {volume} {77}},\ \bibinfo {pages} {134115} (\bibinfo {month}
  {Apr}\ \bibinfo {year} {2008})%
  \bibAnnoteFile{NoStop}{Feng2008}%
\bibitem{Keve1972}%
  \BibitemOpen
  \bibfield{author}{%
  \bibinfo {author} {\bibfnamefont{E.~T.}\ \bibnamefont{Keve}}, \bibinfo
  {author} {\bibfnamefont{K.~L.}\ \bibnamefont{Bye}}, \bibinfo {author}
  {\bibfnamefont{P.~W.}\ \bibnamefont{Whipps}},\ and\ \bibinfo {author}
  {\bibfnamefont{A.~D.}\ \bibnamefont{Annis}},\ }%
  \bibfield{journal}{%
  \Doi{10.1080/00150197108237682}{\bibinfo {journal} {Ferroelectrics}}\ }%
  \textbf{\bibinfo {volume} {3}},\ \bibinfo {pages} {39} (\bibinfo {year}
  {1972})%
  \bibAnnoteFile{NoStop}{Keve1972}%
\bibitem{Robels1993}%
  \BibitemOpen
  \bibfield{author}{%
  \bibinfo {author} {\bibfnamefont{U.}~\bibnamefont{Robels}}\ and\ \bibinfo
  {author} {\bibfnamefont{G.}~\bibnamefont{Arlt}},\ }%
  \bibfield{journal}{%
  \Doi{10.1063/1.352948}{\bibinfo {journal} {Journal of Applied Physics}}\ }%
  \textbf{\bibinfo {volume} {,}},\ \bibinfo {pages} {3454} (\bibinfo {year}
  {1993})%
  \bibAnnoteFile{NoStop}{Robels1993}%
\bibitem{Zhang2008a}%
  \BibitemOpen
  \bibfield{author}{%
  \bibinfo {author} {\bibfnamefont{L.}~\bibnamefont{Zhang}}, \bibinfo {author}
  {\bibfnamefont{E.}~\bibnamefont{Erdem}}, \bibinfo {author}
  {\bibfnamefont{X.}~\bibnamefont{Ren}},\ and\ \bibinfo {author}
  {\bibfnamefont{R.-A.}\ \bibnamefont{Eichel}},\ }%
  \bibfield{journal}{%
  \Doi{10.1063/1.3006327}{\bibinfo {journal} {Applied Physics Letters}}\ }%
  \textbf{\bibinfo {volume} {93}},\ \bibinfo {eid} {202901} (\bibinfo {year}
  {2008})%
  \bibAnnoteFile{NoStop}{Zhang2008a}%
\bibitem{Warren1996}%
  \BibitemOpen
  \bibfield{author}{%
  \bibinfo {author} {\bibfnamefont{W.~L.}\ \bibnamefont{Warren}}, \bibinfo
  {author} {\bibfnamefont{K.}~\bibnamefont{Vanheusden}}, \bibinfo {author}
  {\bibfnamefont{D.}~\bibnamefont{Dimos}}, \bibinfo {author}
  {\bibfnamefont{G.~E.}\ \bibnamefont{Pike}},\ and\ \bibinfo {author}
  {\bibfnamefont{B.~A.}\ \bibnamefont{Tuttle}},\ }%
  \bibfield{journal}{%
  \Doi{10.1111/j.1151-2916.1996.tb08162.x}{\bibinfo {journal} {Journal of the
  American Ceramic Society}}\ }%
  \textbf{\bibinfo {volume} {79}},\ \bibinfo {pages} {536} (\bibinfo {year}
  {1996}),\ ISSN \bibinfo {issn} {1551-2916}%
  \bibAnnoteFile{NoStop}{Warren1996}%
\bibitem{Carl1977}%
  \BibitemOpen
  \bibfield{author}{%
  \bibinfo {author} {\bibfnamefont{K.}~\bibnamefont{Carl}}\ and\ \bibinfo
  {author} {\bibfnamefont{K.~H.}\ \bibnamefont{Hardtl}},\ }%
  \bibfield{journal}{%
  \Doi{10.1080/00150197808236770}{\bibinfo {journal} {Ferroelectrics}}\ }%
  \textbf{\bibinfo {volume} {17}},\ \bibinfo {pages} {473} (\bibinfo {year}
  {1977})%
  \bibAnnoteFile{NoStop}{Carl1977}%
\bibitem{Mestric2004}%
  \BibitemOpen
  \bibfield{author}{%
  \bibinfo {author} {\bibfnamefont{H.}~\bibnamefont{Mestric}}, \bibinfo
  {author} {\bibfnamefont{R.-A.}\ \bibnamefont{Eichel}}, \bibinfo {author}
  {\bibfnamefont{K.-P.}\ \bibnamefont{Dinse}}, \bibinfo {author}
  {\bibfnamefont{A.}~\bibnamefont{Ozarowski}}, \bibinfo {author}
  {\bibfnamefont{J.}~\bibnamefont{van Tol}},\ and\ \bibinfo {author}
  {\bibfnamefont{L.~C.}\ \bibnamefont{Brunel}},\ }%
  \bibfield{journal}{%
  \Doi{10.1063/1.1808477}{\bibinfo {journal} {Journal of Applied Physics}}\ }%
  \textbf{\bibinfo {volume} {96}},\ \bibinfo {pages} {7440} (\bibinfo {year}
  {2004})%
  \bibAnnoteFile{NoStop}{Mestric2004}%
\bibitem{Erhart2007}%
  \BibitemOpen
  \bibfield{author}{%
  \bibinfo {author} {\bibfnamefont{P.}~\bibnamefont{Erhart}}, \bibinfo {author}
  {\bibfnamefont{R.-A.}\ \bibnamefont{Eichel}}, \bibinfo {author}
  {\bibfnamefont{P.}~\bibnamefont{Tr\"askelin}},\ and\ \bibinfo {author}
  {\bibfnamefont{K.}~\bibnamefont{Albe}},\ }%
  \bibfield{journal}{%
  \Doi{10.1103/PhysRevB.76.174116}{\bibinfo {journal} {Phys. Rev. B}}\ }%
  \textbf{\bibinfo {volume} {76}},\ \bibinfo {pages} {174116} (\bibinfo {month}
  {Nov}\ \bibinfo {year} {2007})%
  \bibAnnoteFile{NoStop}{Erhart2007}%
\bibitem{Erhart2013}%
  \BibitemOpen
  \bibfield{author}{%
  \bibinfo {author} {\bibfnamefont{P.}~\bibnamefont{Erhart}}, \bibinfo {author}
  {\bibfnamefont{P.}~\bibnamefont{Tr\"askelin}},\ and\ \bibinfo {author}
  {\bibfnamefont{K.}~\bibnamefont{Albe}},\ }%
  \bibfield{journal}{%
  \Doi{10.1103/PhysRevB.88.024107}{\bibinfo {journal} {Phys. Rev. B}}\ }%
  \textbf{\bibinfo {volume} {88}},\ \bibinfo {pages} {024107} (\bibinfo {month}
  {Jul}\ \bibinfo {year} {2013}),\
  \url{http://link.aps.org/doi/10.1103/PhysRevB.88.024107}%
  \bibAnnoteFile{NoStop}{Erhart2013}%
\bibitem{He2003}%
  \BibitemOpen
  \bibfield{author}{%
  \bibinfo {author} {\bibfnamefont{L.}~\bibnamefont{He}}\ and\ \bibinfo
  {author} {\bibfnamefont{D.}~\bibnamefont{Vanderbilt}},\ }%
  \bibfield{journal}{%
  \Doi{10.1103/PhysRevB.68.134103}{\bibinfo {journal} {Phys. Rev. B}}\ }%
  \textbf{\bibinfo {volume} {68}},\ \bibinfo {pages} {134103} (\bibinfo {month}
  {Oct}\ \bibinfo {year} {2003})%
  \bibAnnoteFile{NoStop}{He2003}%
\bibitem{Gerson1960}%
  \BibitemOpen
  \bibfield{author}{%
  \bibinfo {author} {\bibfnamefont{R.}~\bibnamefont{Gerson}},\ }%
  \bibfield{journal}{%
  \Doi{10.1063/1.1735397}{\bibinfo {journal} {Journal of Applied Physics}}\ }%
  \textbf{\bibinfo {volume} {31}},\ \bibinfo {pages} {188 } (\bibinfo {month}
  {jan}\ \bibinfo {year} {1960}),\ ISSN \bibinfo {issn} {0021-8979}%
  \bibAnnoteFile{NoStop}{Gerson1960}%
\bibitem{Eyraud2006}%
  \BibitemOpen
  \bibfield{author}{%
  \bibinfo {author} {\bibfnamefont{L.}~\bibnamefont{Eyraud}}, \bibinfo {author}
  {\bibfnamefont{B.}~\bibnamefont{Guiffard}}, \bibinfo {author}
  {\bibfnamefont{L.}~\bibnamefont{Lebrun}},\ and\ \bibinfo {author}
  {\bibfnamefont{D.}~\bibnamefont{Guyomar}},\ }%
  \bibfield{journal}{%
  \Doi{10.1080/00150190600605510}{\bibinfo {journal} {Ferroelectrics}}\ }%
  \textbf{\bibinfo {volume} {330}},\ \bibinfo {pages} {51} (\bibinfo {year}
  {2006})%
  \bibAnnoteFile{NoStop}{Eyraud2006}%
\bibitem{Safari2008}%
  \BibitemOpen
  \bibfield{author}{%
  \bibinfo {author} {\bibfnamefont{A.}~\bibnamefont{Safari}}\ and\ \bibinfo
  {author} {\bibfnamefont{E.~K.}\ \bibnamefont{Akdo{\u{g}}an}},\ }%
  \emph{\bibinfo {title} {Piezoelectric and acoustic materials for transducer
  applications}}\ (\bibinfo {publisher} {Springer},\ \bibinfo {year} {2008})%
  \bibAnnoteFile{NoStop}{Safari2008}%
\bibitem{Whatmore1986}%
  \BibitemOpen
  \bibfield{author}{%
  \bibinfo {author} {\bibfnamefont{R.}~\bibnamefont{Whatmore}},\ }%
  \bibfield{journal}{%
  \bibinfo {journal} {Reports on progress in physics}\ }%
  \textbf{\bibinfo {volume} {49}},\ \bibinfo {pages} {1335} (\bibinfo {year}
  {1986})%
  \bibAnnoteFile{NoStop}{Whatmore1986}%
\bibitem{Takeuchi1982}%
  \BibitemOpen
  \bibfield{author}{%
  \bibinfo {author} {\bibfnamefont{H.}~\bibnamefont{Takeuchi}}, \bibinfo
  {author} {\bibfnamefont{S.}~\bibnamefont{Jyomura}}, \bibinfo {author}
  {\bibfnamefont{E.}~\bibnamefont{Yamamoto}},\ and\ \bibinfo {author}
  {\bibfnamefont{Y.}~\bibnamefont{Ito}},\ }%
  \bibfield{journal}{%
  \bibinfo {journal} {The Journal of the Acoustical Society of America}\ }%
  \textbf{\bibinfo {volume} {72}},\ \bibinfo {pages} {1114} (\bibinfo {year}
  {1982})%
  \bibAnnoteFile{NoStop}{Takeuchi1982}%
\bibitem{Meyer2002}%
  \BibitemOpen
  \bibfield{author}{%
  \bibinfo {author} {\bibfnamefont{B.}~\bibnamefont{Meyer}}\ and\ \bibinfo
  {author} {\bibfnamefont{D.}~\bibnamefont{Vanderbilt}},\ }%
  \bibfield{journal}{%
  \Doi{10.1103/PhysRevB.65.104111}{\bibinfo {journal} {Phys. Rev. B}}\ }%
  \textbf{\bibinfo {volume} {65}},\ \bibinfo {pages} {104111} (\bibinfo {month}
  {Mar}\ \bibinfo {year} {2002})%
  \bibAnnoteFile{NoStop}{Meyer2002}%
\bibitem{Giannozzi2009}%
  \BibitemOpen
  \bibfield{author}{%
  \bibinfo {author} {\bibfnamefont{P.}~\bibnamefont{Giannozzi}}, \bibinfo
  {author} {\bibfnamefont{S.}~\bibnamefont{Baroni}}, \bibinfo {author}
  {\bibfnamefont{N.}~\bibnamefont{Bonini}}, \bibinfo {author}
  {\bibfnamefont{M.}~\bibnamefont{Calandra}}, \bibinfo {author}
  {\bibfnamefont{R.}~\bibnamefont{Car}}, \bibinfo {author}
  {\bibfnamefont{C.}~\bibnamefont{Cavazzoni}}, \bibinfo {author}
  {\bibfnamefont{D.}~\bibnamefont{Ceresoli}}, \bibinfo {author}
  {\bibfnamefont{G.~L.}\ \bibnamefont{Chiarotti}}, \bibinfo {author}
  {\bibfnamefont{M.}~\bibnamefont{Cococcioni}}, \bibinfo {author}
  {\bibfnamefont{I.}~\bibnamefont{Dabo}}, \emph{et~al.},\ }%
  \bibfield{journal}{%
  \bibinfo {journal} {Journal of Physics: Condensed Matter}\ }%
  \textbf{\bibinfo {volume} {21}},\ \bibinfo {pages} {395502} (\bibinfo {year}
  {2009})%
  \bibAnnoteFile{NoStop}{Giannozzi2009}%
\bibitem{Saghi-Szabo1999}%
  \BibitemOpen
  \bibfield{author}{%
  \bibinfo {author} {\bibfnamefont{G.}~\bibnamefont{S\'aghi-Szab\'o}}, \bibinfo
  {author} {\bibfnamefont{R.~E.}\ \bibnamefont{Cohen}},\ and\ \bibinfo {author}
  {\bibfnamefont{H.}~\bibnamefont{Krakauer}},\ }%
  \bibfield{journal}{%
  \Doi{10.1103/PhysRevB.59.12771}{\bibinfo {journal} {Phys. Rev. B}}\ }%
  \textbf{\bibinfo {volume} {59}},\ \bibinfo {pages} {12771} (\bibinfo {month}
  {May}\ \bibinfo {year} {1999}),\
  \url{http://link.aps.org/doi/10.1103/PhysRevB.59.12771}%
  \bibAnnoteFile{NoStop}{Saghi-Szabo1999}%
\bibitem{Yao2011}%
  \BibitemOpen
  \bibfield{author}{%
  \bibinfo {author} {\bibfnamefont{Y.}~\bibnamefont{Yao}}\ and\ \bibinfo
  {author} {\bibfnamefont{H.}~\bibnamefont{Fu}},\ }%
  \bibfield{journal}{%
  \Doi{10.1103/PhysRevB.84.064112}{\bibinfo {journal} {Phys. Rev. B}}\ }%
  \textbf{\bibinfo {volume} {84}},\ \bibinfo {pages} {064112} (\bibinfo {month}
  {Aug}\ \bibinfo {year} {2011})%
  \bibAnnoteFile{NoStop}{Yao2011}%
\bibitem{Nowick1965}%
  \BibitemOpen
  \bibfield{author}{%
  \bibinfo {author} {\bibfnamefont{A.}~\bibnamefont{Nowick}}\ and\ \bibinfo
  {author} {\bibfnamefont{W.}~\bibnamefont{Heller}},\ }%
  \bibfield{journal}{%
  \Doi{10.1080/00018736500101021}{\bibinfo {journal} {Advances in Physics}}\ }%
  \textbf{\bibinfo {volume} {14}},\ \bibinfo {pages} {101} (\bibinfo {year}
  {1965})%
  \bibAnnoteFile{NoStop}{Nowick1965}%
\bibitem{Morozov2008}%
  \BibitemOpen
  \bibfield{author}{%
  \bibinfo {author} {\bibfnamefont{M.~I.}\ \bibnamefont{Morozov}}\ and\
  \bibinfo {author} {\bibfnamefont{D.}~\bibnamefont{Damjanovic}},\ }%
  \bibfield{journal}{%
  \Doi{10.1063/1.2963704}{\bibinfo {journal} {Journal of Applied Physics}}\ }%
  \textbf{\bibinfo {volume} {104}},\ \bibinfo {eid} {034107} (\bibinfo {year}
  {2008})%
  \bibAnnoteFile{NoStop}{Morozov2008}%
\bibitem{Henkelman2000}%
  \BibitemOpen
  \bibfield{author}{%
  \bibinfo {author} {\bibfnamefont{G.}~\bibnamefont{Henkelman}}, \bibinfo
  {author} {\bibfnamefont{B.~P.}\ \bibnamefont{Uberuaga}},\ and\ \bibinfo
  {author} {\bibfnamefont{H.}~\bibnamefont{J{\'o}nsson}},\ }%
  \bibfield{journal}{%
  \bibinfo {journal} {The Journal of Chemical Physics}\ }%
  \textbf{\bibinfo {volume} {113}},\ \bibinfo {pages} {9901} (\bibinfo {year}
  {2000})%
  \bibAnnoteFile{NoStop}{Henkelman2000}%
\bibitem{Dai1991}%
  \BibitemOpen
  \bibfield{author}{%
  \bibinfo {author} {\bibfnamefont{G.~H.}\ \bibnamefont{Dai}}, \bibinfo
  {author} {\bibfnamefont{P.~W.}\ \bibnamefont{Lu}}, \bibinfo {author}
  {\bibfnamefont{X.~Y.}\ \bibnamefont{Huang}}, \bibinfo {author}
  {\bibfnamefont{Q.~S.}\ \bibnamefont{Liu}},\ and\ \bibinfo {author}
  {\bibfnamefont{W.~R.}\ \bibnamefont{Xue}},\ }%
  \bibfield{journal}{%
  \bibinfo {journal} {Journal of Materials Science: Materials in Electronics}\
  }%
  \textbf{\bibinfo {volume} {2}},\ \bibinfo {pages} {164} (\bibinfo {year}
  {1991}),\ ISSN \bibinfo {issn} {0957-4522}%
  \bibAnnoteFile{NoStop}{Dai1991}%
\bibitem{Cockayne2004}%
  \BibitemOpen
  \bibfield{author}{%
  \bibinfo {author} {\bibfnamefont{E.}~\bibnamefont{Cockayne}}\ and\ \bibinfo
  {author} {\bibfnamefont{B.~P.}\ \bibnamefont{Burton}},\ }%
  \bibfield{journal}{%
  \Doi{10.1103/PhysRevB.69.144116}{\bibinfo {journal} {Phys. Rev. B}}\ }%
  \textbf{\bibinfo {volume} {69}},\ \bibinfo {pages} {144116} (\bibinfo {month}
  {Apr}\ \bibinfo {year} {2004})%
  \bibAnnoteFile{NoStop}{Cockayne2004}%
\bibitem{Poykko2000}%
  \BibitemOpen
  \bibfield{author}{%
  \bibinfo {author} {\bibfnamefont{S.}~\bibnamefont{Poykko}}\ and\ \bibinfo
  {author} {\bibfnamefont{D.~J.}\ \bibnamefont{Chadi}},\ }%
  \bibfield{journal}{%
  \Doi{10.1063/1.125800}{\bibinfo {journal} {Applied Physics Letters}}\ }%
  \textbf{\bibinfo {volume} {76}},\ \bibinfo {pages} {499} (\bibinfo {year}
  {2000})%
  \bibAnnoteFile{NoStop}{Poykko2000}%
\bibitem{Resta1994}%
  \BibitemOpen
  \bibfield{author}{%
  \bibinfo {author} {\bibfnamefont{R.}~\bibnamefont{Resta}},\ }%
  \bibfield{journal}{%
  \Doi{10.1103/RevModPhys.66.899}{\bibinfo {journal} {Rev. Mod. Phys.}}\ }%
  \textbf{\bibinfo {volume} {66}},\ \bibinfo {pages} {899} (\bibinfo {month}
  {Jul}\ \bibinfo {year} {1994}),\
  \url{http://link.aps.org/doi/10.1103/RevModPhys.66.899}%
  \bibAnnoteFile{NoStop}{Resta1994}%
\bibitem{Mackie2010}%
  \BibitemOpen
  \bibfield{author}{%
  \bibinfo {author} {\bibfnamefont{R.~A.}\ \bibnamefont{Mackie}}, \bibinfo
  {author} {\bibfnamefont{A.}~\bibnamefont{Pel\'aiz-Barranco}},\ and\ \bibinfo
  {author} {\bibfnamefont{D.~J.}\ \bibnamefont{Keeble}},\ }%
  \bibfield{journal}{%
  \Doi{10.1103/PhysRevB.82.024113}{\bibinfo {journal} {Phys. Rev. B}}\ }%
  \textbf{\bibinfo {volume} {82}},\ \bibinfo {pages} {024113} (\bibinfo {month}
  {Jul}\ \bibinfo {year} {2010})%
  \bibAnnoteFile{NoStop}{Mackie2010}%
\bibitem{Eichel2006}%
  \BibitemOpen
  \bibfield{author}{%
  \bibinfo {author} {\bibfnamefont{R.-A.}\ \bibnamefont{Eichel}}, \bibinfo
  {author} {\bibfnamefont{H.}~\bibnamefont{Mestric}}, \bibinfo {author}
  {\bibfnamefont{H.}~\bibnamefont{Kungl}},\ and\ \bibinfo {author}
  {\bibfnamefont{M.~J.}\ \bibnamefont{Hoffmann}},\ }%
  \bibfield{journal}{%
  \Doi{10.1063/1.2185258}{\bibinfo {journal} {Applied Physics Letters}}\ }%
  \textbf{\bibinfo {volume} {88}},\ \bibinfo {eid} {122506} (\bibinfo {year}
  {2006})%
  \bibAnnoteFile{NoStop}{Eichel2006}%
\bibitem{Shimada2013}%
  \BibitemOpen
  \bibfield{author}{%
  \bibinfo {author} {\bibfnamefont{T.}~\bibnamefont{Shimada}}, \bibinfo
  {author} {\bibfnamefont{T.}~\bibnamefont{Ueda}}, \bibinfo {author}
  {\bibfnamefont{J.}~\bibnamefont{Wang}},\ and\ \bibinfo {author}
  {\bibfnamefont{T.}~\bibnamefont{Kitamura}},\ }%
  \bibfield{journal}{%
  \Doi{10.1103/PhysRevB.87.174111}{\bibinfo {journal} {Phys. Rev. B}}\ }%
  \textbf{\bibinfo {volume} {87}},\ \bibinfo {pages} {174111} (\bibinfo {month}
  {May}\ \bibinfo {year} {2013}),\
  \url{http://link.aps.org/doi/10.1103/PhysRevB.87.174111}%
  \bibAnnoteFile{NoStop}{Shimada2013}%
\bibitem{Jin2009}%
  \BibitemOpen
  \bibfield{author}{%
  \bibinfo {author} {\bibfnamefont{L.}~\bibnamefont{Jin}}, \bibinfo {author}
  {\bibfnamefont{Z.}~\bibnamefont{He}},\ and\ \bibinfo {author}
  {\bibfnamefont{D.}~\bibnamefont{Damjanovic}},\ }%
  \bibfield{journal}{%
  \bibinfo {journal} {Applied Physics Letters}\ }%
  \textbf{\bibinfo {volume} {95}},\ \bibinfo {pages} {012905} (\bibinfo {year}
  {2009})%
  \bibAnnoteFile{NoStop}{Jin2009}%
\bibitem{Maksymovych2012}%
  \BibitemOpen
  \bibfield{author}{%
  \bibinfo {author} {\bibfnamefont{P.}~\bibnamefont{Maksymovych}}, \bibinfo
  {author} {\bibfnamefont{A.~N.}\ \bibnamefont{Morozovska}}, \bibinfo {author}
  {\bibfnamefont{P.}~\bibnamefont{Yu}}, \bibinfo {author}
  {\bibfnamefont{E.~A.}\ \bibnamefont{Eliseev}}, \bibinfo {author}
  {\bibfnamefont{Y.-H.}\ \bibnamefont{Chu}}, \bibinfo {author}
  {\bibfnamefont{R.}~\bibnamefont{Ramesh}}, \bibinfo {author}
  {\bibfnamefont{A.~P.}\ \bibnamefont{Baddorf}},\ and\ \bibinfo {author}
  {\bibfnamefont{S.~V.}\ \bibnamefont{Kalinin}},\ }%
  \bibfield{journal}{%
  \bibinfo {journal} {Nano Letters}\ }%
  \textbf{\bibinfo {volume} {12}},\ \bibinfo {pages} {209} (\bibinfo {year}
  {2012})%
  \bibAnnoteFile{NoStop}{Maksymovych2012}%
\bibitem{Zhang2006a}%
  \BibitemOpen
  \bibfield{author}{%
  \bibinfo {author} {\bibfnamefont{Z.}~\bibnamefont{Zhang}}, \bibinfo {author}
  {\bibfnamefont{L.}~\bibnamefont{Lu}}, \bibinfo {author}
  {\bibfnamefont{C.}~\bibnamefont{Shu}},\ and\ \bibinfo {author}
  {\bibfnamefont{P.}~\bibnamefont{Wu}},\ }%
  \bibfield{journal}{%
  \Doi{10.1063/1.2362993}{\bibinfo {journal} {Applied Physics Letters}}\ }%
  \textbf{\bibinfo {volume} {89}},\ \bibinfo {eid} {152909} (\bibinfo {year}
  {2006})%
  \bibAnnoteFile{NoStop}{Zhang2006a}%
\end{thebibliography}%
\end{document}